\documentstyle[12pt,amssymbols,epsf]{article}
\newcommand{\twm}{two-charge model}
\newcommand{\beqn}{\begin{equation}}
\newcommand{\eeqn}{\end{equation}}
\newcommand{\beqa}{\begin{eqnarray}}
\newcommand{\eeqa}{\end{eqnarray}}
\newcommand{\MC}{Monte Carlo}

\newcommand{\RGF}{renormalization group flow}
\newcommand{\eqs}{equation of state}
\newcommand{\eqss}{equations of state}
\newcommand{\chiu}{\langle\bar{\chi}_{u}\chi_{u}\rangle}
\newcommand{\chid}{\langle\bar{\chi}_{d}\chi_{d}\rangle}
\newcommand{\mpd}{$m_{\pi_d}$}
\begin{document}
\begin{titlepage}

\begin{flushright}
HLRZ 66/95\\
GUTPA/95/10/05\\
November 1995
\end{flushright}

\vspace*{3mm}

\begin{center}
{\Huge Non-compact Lattice QED \\
with Two Charges: \\
Phase Diagram and
Renormalization Group Flow
}\\[12mm]

{ \large \bf A.~Ali Khan} \\
Department of Physics and Astronomy, University of Glasgow, \\
Glasgow G12~8QQ, UK

\end{center}
\vspace{3mm}
\begin{abstract}
\noindent The phase diagram of non-compact lattice QED in four dimensions with
staggered
fermions of charges 1 and $-1/2$ is investigated.
The renormalized charges are  determined and found to be in
agreement with  perturbation theory. This is an indication that there
is no  continuum limit with non-vanishing renormalized gauge coupling,
and that the theory has a validity bound for every finite value of the
renormalized coupling.
 The renormalization group flow
of the charges is investigated and an estimate for the validity
 bound  as a function of the cut-off
is obtained. Generalizing this estimate to all fermions in the Standard Model,
it is found  that a cut-off at the Planck scale
implies that $\alpha_R$ has to be less
than $1/80$. Due to spontaneous chiral  symmetry breaking,
strongly bound fermion-antifermion composite states are generated. Their
spectrum is  discussed.
\end{abstract}
\thispagestyle{empty}
\end{titlepage}
\newpage
\section{Introduction}
 Interest in non-perturbative investigations of QED has a long history.
{}From perturbation theory there are indications (Landau
pole~\cite{landau}) that the cut-off can only be removed from the theory
if the renormalized charge vanishes. Also, it is a yet unsolved fundamental
question~\cite{feynman85}, why the fine structure constant has the
value $\alpha \simeq 1/137$.  At strong coupling, QED exhibits
a chiral symmetry breaking phase transition of 2nd
order~\cite{miransky,bartho84,salm90,kondo91a},
where tightly bound
fermion-antifermion pairs are generated. There one could expect a
deviation from the charge screening behaviour known from perturbative QED.
The phase transition makes strongly coupled $U(1)$ gauge theories interesting
also for applications in technicolour theories because of large anomalous
dimensions of the operator $\bar{\psi}\psi$~\cite{anom-dim}.
For several years, the possibility of a non-trivial continuum limit, or a
non-trivial ultraviolet-stable fixed point of the Callan-Symanzik
$\beta$ function, has been investigated in the
non-compact formulation of QED on the lattice (see for
example~\cite{kogu88}--\cite{azco92}).

Studying non-compact lattice QED with dynamical
fermions,  some groups find
non-trivial scaling behaviour~\cite{Kocic93,Kocic94,zaragoza_new}, others
find their critical exponents to indicate triviality~\cite{boot89,juel94}.
The QED
$\beta$ function  has been studied non--perturbatively
on the lattice~\cite{schi92,horo91}, and using
Schwinger-Dyson equations~\cite{rakow91}. It
turns out to be in agreement with  perturbation theory
and to show no indications of a non-trivial fixed point. If QED is
trivial in the limit of infinite cut-off, there is a maximal cut-off
corresponding to every finite value of the renormalized charge. One
is interested in the size of this validity bound.
A rough estimate of it has
been made in a lattice study of QED with  one charge~\cite{schi92}.

In nature one finds several differently charged types of fermions. In the
presence of several charges, new non-perturbative phenomena may arise which
could affect the phase structure.  An important question is whether
chiral symmetry breaking
sets in at the same value of the bare coupling for all fermions or whether some
fermion species can be massless and others in the chirally broken phase at the
same time.  The phase diagram of such a system has to be
investigated with respect to a physically interesting continuum limit.
In QED with one charge,
neutral pointlike Goldstone bosons and scalar particles are generated due to
the chiral symmetry breaking phase transition.  One could expect that in the
\twm\  there is a larger number of neutral pointlike
bound states which could in principle carry colour charge and have an effect
on the $\beta$ function of QCD.  Moreover, electrically
charged bound states may appear
which could change the behaviour of the QED  $\beta$ function and push the
validity bound towards much lower energies.  It is of interest to
see whether charge renormalization is in agreement with
perturbation theory and which states give relevant contributions.

To  investigate non-perturbative phenomena in the coupling
between different species of
fermions,  a model with two species of
staggered fermions with a ratio of their charges of $-1/2$ (`two-charge
model') was studied,  comparable to $u$ and $d$ quarks whose strong and weak
interactions are switched off.

In section~\ref{sec:phase} of this paper a description of the model and
details  of the simulation are given. The phase diagram,
obtained from the chiral condensates, and the scaling behaviour are discussed.
The charged sector of the model
 is presented in section~\ref{sec:charge}. The renormalized charges
are determined non-perturbatively using current-photon correlation functions
and are compared with
renormalized lattice perturbation theory.  There is  good agreement which
leads to the conclusion that the model is trivial. Using perturbation theory,
the renormalization group flow of the charges is determined, which leads
to an estimate of the validity bound  resulting from triviality.
Spectrum and renormalization group flows of fermion-antifermion composite
states are discussed in section~\ref{sec:comp}.
\section{\label{sec:phase} Phase diagram and scaling behaviour}
\subsection{Action and Simulation Details}
The `two-charge model' contains a non-compact gauge field with the action
\beqn
S_{g} = \frac{\beta}{2} \sum_{x} \sum_{\mu < \nu} F_{\mu
  \nu}^{2}(x),\label{eq:eich}
\eeqn
\beqn
F_{\mu \nu} (x) = \Delta_{\mu}A_{\nu}(x) - \Delta_{\nu}A_{\mu}(x),
\eeqn
and two sets with each four flavours
of staggered fermions which couple to the gauge fields with
couplings 1 and --1/2, corresponding to $u$ and $d$ quarks with only
electromagnetic interactions:
\beqn
S_{f}  =  \sum_{x,y}\left\{ \bar{\chi}_{u}(x) M_{u,xy}
\chi_{u}(y)
 +  \bar{\chi}_{d}(x)M_{d,xy} \chi_{d}(y)\right\}. \label{eq:fermwir}
\eeqn
The coupling $\beta$ is related to the bare electric charges $e_k$ by
$\beta = 1/(c_ke_k)^2$, $k = 1,2$.
The lattice spacing $a$ is set to 1. The fermion matrices are given by
\beqa
M_{k,x,y}&= &D_{k,xy}+m_{k}\delta_{xy} \nonumber \\
D_{k,xy}(x) & = & \frac{1}{2}\sum_{\mu} \eta_{\mu}(x)
 \left\{ e^{c_{k}iA_{\mu}(x)}\delta_{y,x+ \hat{\mu}} -
 e^{-c_{k}iA_{\mu}(y)}\delta_{y,x- \hat{\mu}} \right\};  \\
k = u, d & ; & c_{u} = 1,\; c_{d} = -1/2 \nonumber .
\eeqa
In the limit $m_d \rightarrow \infty$ this model goes over into non-compact
QED with one charge~\cite{kogu88,schi90}.
For the gauge fields, periodic boundary conditions in all four
directions were chosen, for the fermions periodic spatial and antiperiodic
temporal  boundary conditions. The simulations were
performed on lattices of size $8^3\times 12$. From simulations with one charge,
one expects that for the chosen values of $\beta$, $m_u$ and $m_d$ finite
size effects are small~\cite{schi92}. For each
simulated point $(\beta,m_u,m_d)$
$O(1000)$ configurations in equilibrium were generated using a Hybrid Monte
Carlo algorithm.
Every fifth was stored for spectrum and charge calculations.

Staggered fermions are a useful choice for studying chiral symmetry
properties at finite lattice spacing.
The action (\ref{eq:fermwir}) of the \twm\ has  for the $k$th species of
fermions a chiral $U(1)_{V}\times U(1)_{A}$ symmetry, if $m_k = 0$. The
 order parameters are the chiral condensates
\begin{equation}
\sigma_k \equiv \langle \bar{\chi}_k\chi_k\rangle  = -\langle Tr M^{-1}_k
\rangle,
\end{equation}
which are computed using a stochastic
estimator~\cite{Bit89}.
Simulation results for the chiral condensates are shown in tables~\ref{table1}
and~\ref{table2}.
\subsection{Determination of the critical points}
\begin{figure}[pthb]
\centerline{\epsfxsize=7.5cm
\epsfbox{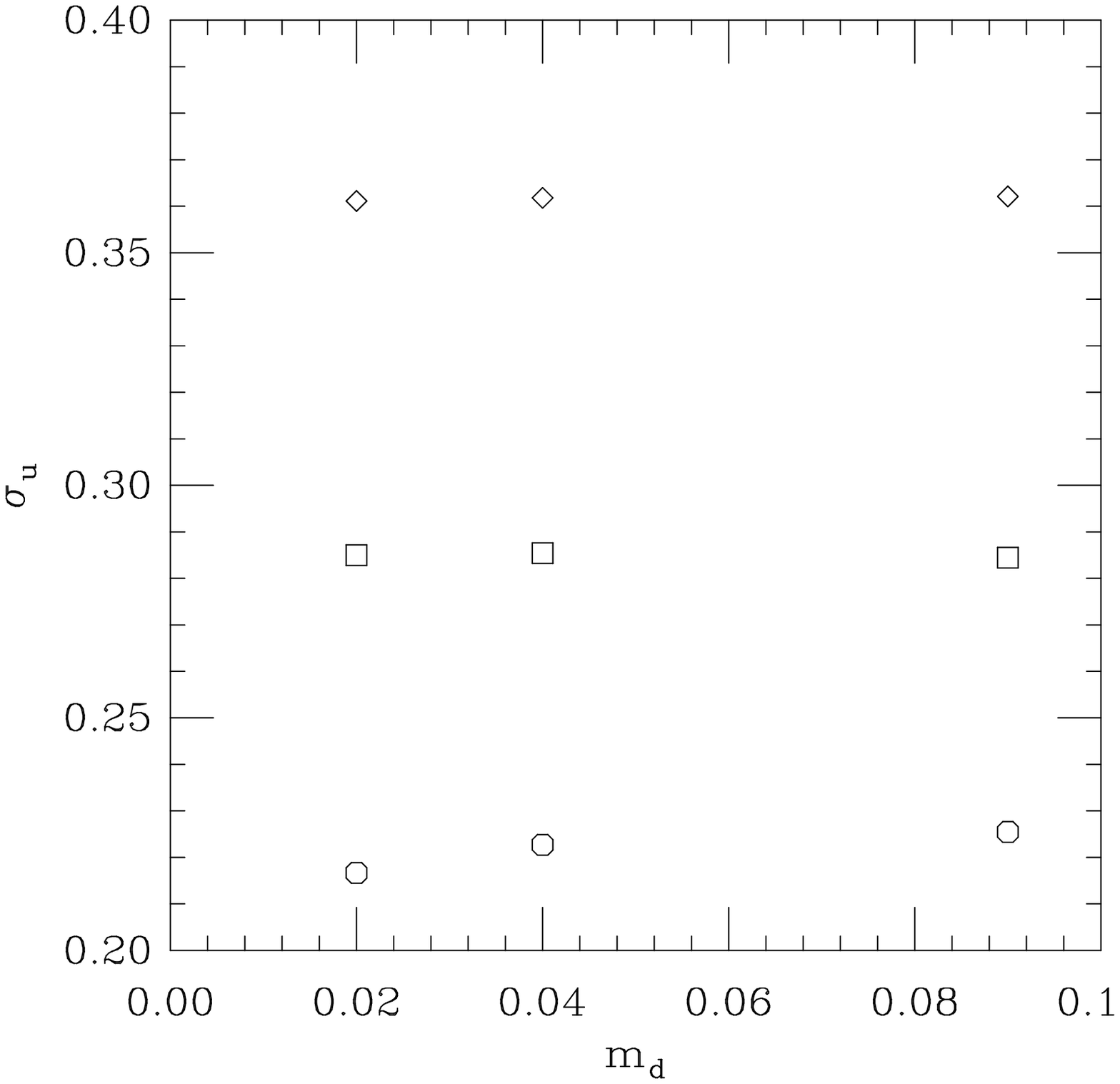}
\epsfxsize=7.5cm
\epsfbox{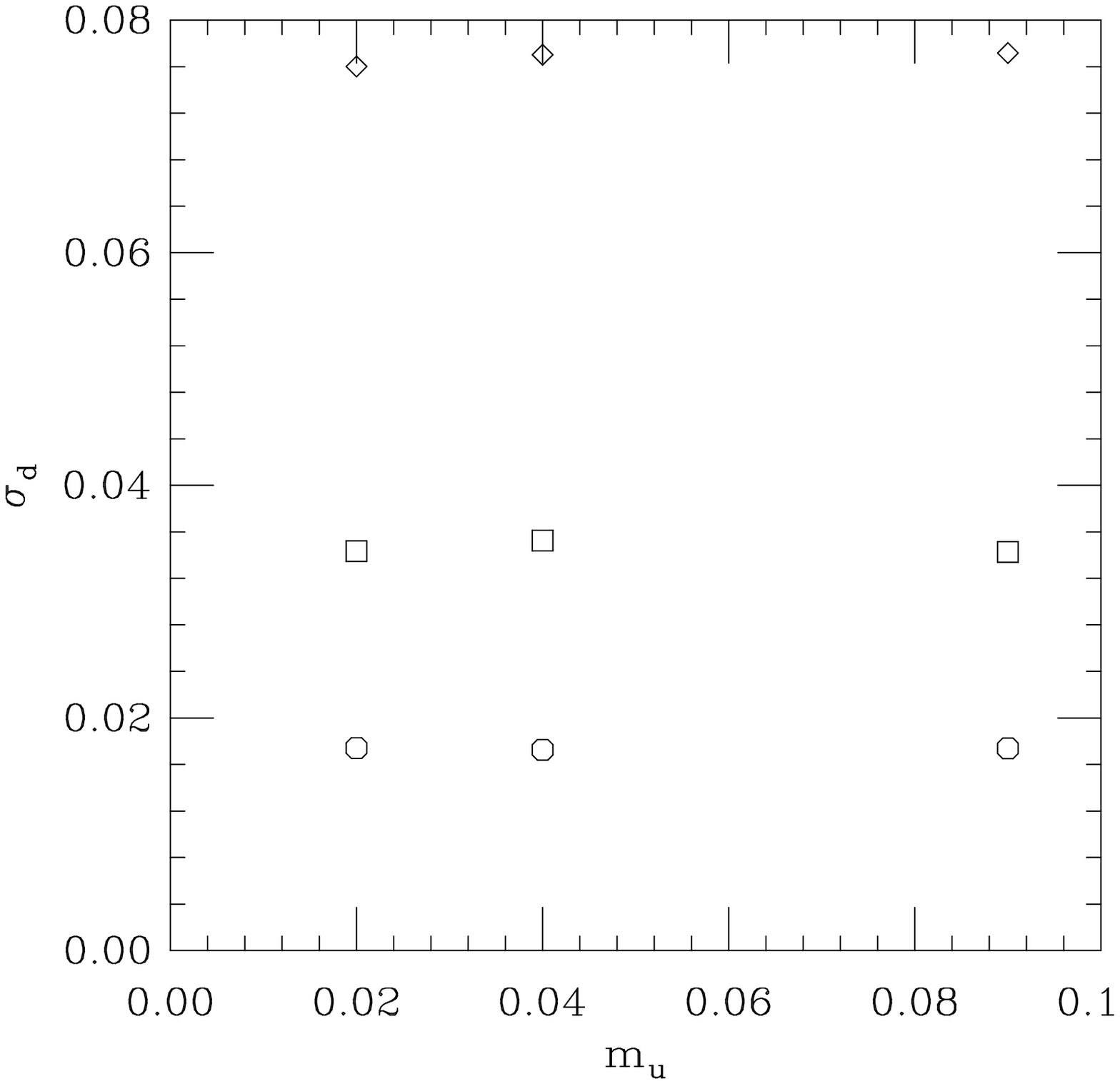}}
  \caption{Dependence of $\sigma_u$ on $m\equiv m_d$ (left) and of $\sigma_d$
on
$m \equiv m_u$ (right). Circles denote $m$ = 0.02, squares $m$ = 0.04 and
diamonds $m$ = 0.09. $\beta = 0.18$. Statistical errors are smaller than the
symbols.}
 \label{fig:sumd}
\end{figure}

In non-compact QED with one set of staggered fermions the chiral condensate
$\sigma $
is consistent with a mean field like equation of state  with logarithmic
corrections motivated from a
linear $\sigma$ model~\cite{schi92,brezin76}.
The parameters in the \eqs\ are expanded in a power series in the
reduced coupling $(1-\beta/\beta_c)$, where $\beta_c$ denotes the critical
coupling.
The logarithmic corrections are only expected to become important very close to
the critical point due to renormalization effects. It is expected that they
become relevant also here if one goes closer to the critical point.
{}From the results in tables~\ref{table1} and~\ref{table2},  and as illustrated
for $\beta = 0.18$ in figure~\ref{fig:sumd},  it appears that
the chiral condensates are fairly
independent of the other fermion's bare mass.
Thus to determine the critical points in the \twm\, an Ansatz with two
uncoupled mean field  like equations of state with logarithmic corrections
 for each fermion is used:
\begin{equation}
m_{k}=2\tau_{k}\frac{\sigma_{k}}{\ln^{p_{k}}|\sigma_{k}^{-1}|}
+4\theta_{k}\frac{\sigma_{k}^{3}}{\ln|\sigma_{k}^{-1}|},\;\; k=u,d
 \label{eq:state}.
\end{equation}
For the determination of the \RGF\ it is desirable to know the chiral
condensates as a function of $(\beta,m_u,m_d)$ in the whole parameter space.
It turns out to be possible to approximate
 the chiral condensates for all $\beta$  with \eqss\ (\ref{eq:state}),
using the following expansion of the couplings:
\begin{eqnarray}
 \tau_{u}/\theta_{u} & = & \tau_{u}^{(1)}(1-\beta/
\beta_{cu}) +\tau_{u}^{(3)}(1 - \beta/\beta_{cu})^{3} , \nonumber \\
1/\theta_{u} & = & \theta_{u}^{(0)}+\theta_{u}^{(1)}
(1 - \beta/\beta_{cu})+\theta_{u}^{(3)}(1 - \beta/\beta_{cu})^{3},\label{eq:cu}
\end{eqnarray}
and
\begin{eqnarray}
 \tau_{d}/\theta_{d} & = & \tau_{d}^{(1)}(1-\beta/
\beta_{cd})  , \nonumber \\
1/\theta_{d} & = & \theta_{d}^{(0)}+\theta_{d}^{(1)}
(1 - \beta/\beta_{cd}).
\end{eqnarray}
\begin{figure}[bhtp]
\vspace{-5.6cm}
\epsfysize=18cm
\centerline{\epsfbox{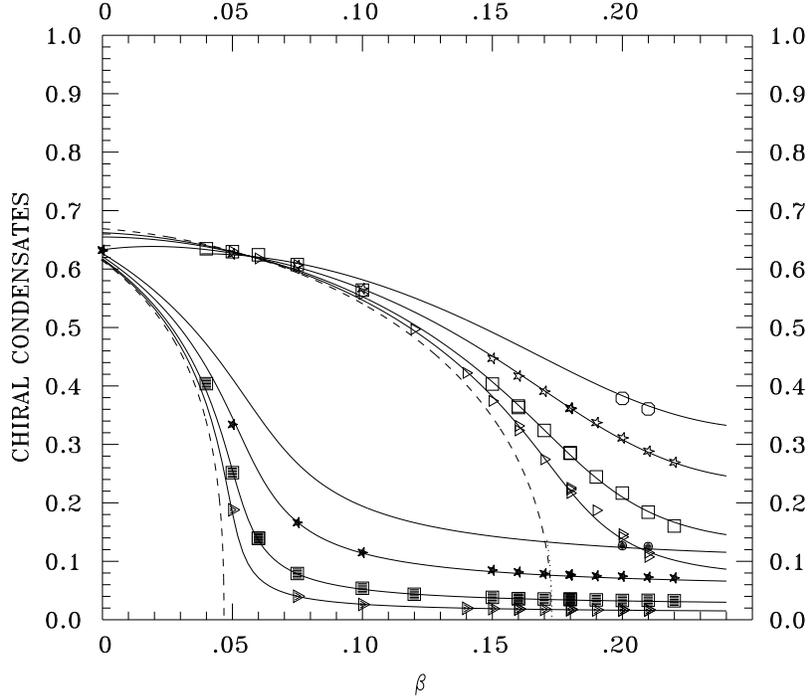}}
\vspace{-3.5cm}
  \caption{Chiral condensates (open symbols: $\sigma_u$, filled symbols:
$\sigma_d$) as a function of $\beta$ with a mean field
fit with logarithmic corrections. Triangles correspond to $m_{k}$ = 0.02,
squares to
 $m_{k}$ = 0.04, asterisks to $m_{k}$ = 0.09 and circles to $m_{k}$
= 0.16, $k = u,d$.
 The solid lines show the result of the fits and the dashed lines
solutions of the equations of state with $m_{k}$ set to 0. }
 \label{fig:chis}
\end{figure}
$\beta_{ck}$, $\tau_{k}^{(1,3)}$, $\theta_{k}^{(0,1,3)}$
and $p_{k}$ are fit parameters.
 Including all results  for $\sigma_u$ simulated at
 $m_{u}\leq 0.09$, one obtains $\beta_{cu}$ = 0.173.  Values of the fit
parameters and the fit errors are listed in table~\ref{tab:mf-logs}.
For a fit of $\sigma_{d}$  all results  with  $\beta\leq 0.1$ and
$m_{d}\leq 0.09$ have been used, and the critical coupling is
 $\beta_{cd}=0.047$. From
varying the range of the chiral condensates included in the fit,
 one estimates the error on $\beta_{cu}$ and $\beta_{cd}$ to be approximately
0.001, which is larger than the fit error.
Without including logarithmic corrections,
$\beta_{cu}$ comes out to be $0.183(1)$ and $\beta_{cd}$ to be
$0.049(1)$~\cite{alikhan92}. The  cubic term in eq.~(\ref{eq:cu}) is
included to obtain an approximate description of $\sigma_u$ also in the
region where $\beta \sim \beta_{cd}$. It has been
checked that with a fit in the range $0.16 \leq \beta \leq 0.22$, including
only linear terms in the reduced coupling, one obtains the same result for
$\beta_{cu}$ within errors.
As seen in figure~\ref{fig:chis}, eqs.~(\ref{eq:state}) give a good
description for
the chiral condensates in the range of couplings $\beta \geq 0.05$.
Two or three of the
same symbols lying on top of each other at the $\beta$
values 0.06, 0.16, 0.18, 0.20 and 0.21 corresponds to simulations at
various values of $m_u$ at a fixed value of $m_d$ or vice versa.
One
further notices that in the regions where the $d$ fermion undergoes a
transition, the $u$ fermion is that far in the broken region that
$\sigma_u$ is practically independent of $m_u$ (as well as of $m_d$).
\subsection{Renormalized fermion masses}
The next step in the investigation of the critical behaviour  is the
determination of the renormalized fermion masses or inverse
fermionic correlation lengths.
Because the fermions are charged, their correlation functions are gauge
dependent. For the calculation of their expectation value, a technique as
described
in references~\cite{goeckeler91,schi92} is used. First, Landau gauge is fixed
by imposing the  following  gauge fixing condition:
\begin{equation}
\sum_{\mu} \bar{\Delta}_{\mu}\, A_{\mu}(x) = 0,
\end{equation}
where $\bar{\Delta}_{\mu}$ denotes the backward derivative on the lattice.
An additional gauge-like degree of freedom is the invariance of the action
under the local transformation
\begin{eqnarray}
A_{\mu}(x) & \to & A_{\mu}(x) + \Delta_{\mu}\, \alpha (x),
\nonumber \\
\chi(x) & \to & \mbox{e}^{- i c_k \alpha (x)} \chi_k (x),
\nonumber \\
\bar{\chi}_k(x) & \to & \mbox{e}^{{ i}c_{k} \alpha (x)}
\bar{\chi}_k (x),
\end{eqnarray}
with
\begin{equation}
\alpha(x) = \sum_{\mu} \frac{4\pi}{L_{\mu}} n_{\mu} x_{\mu} \; ,
 n_{\mu} \in {\Bbb Z},
\end{equation}
where $L_\mu$ is the lattice extent in the $\mu$ direction.
\begin{figure}[p]
\leavevmode
\vspace{-5.5cm}
\centerline{
\epsfysize=14cm
\epsfbox{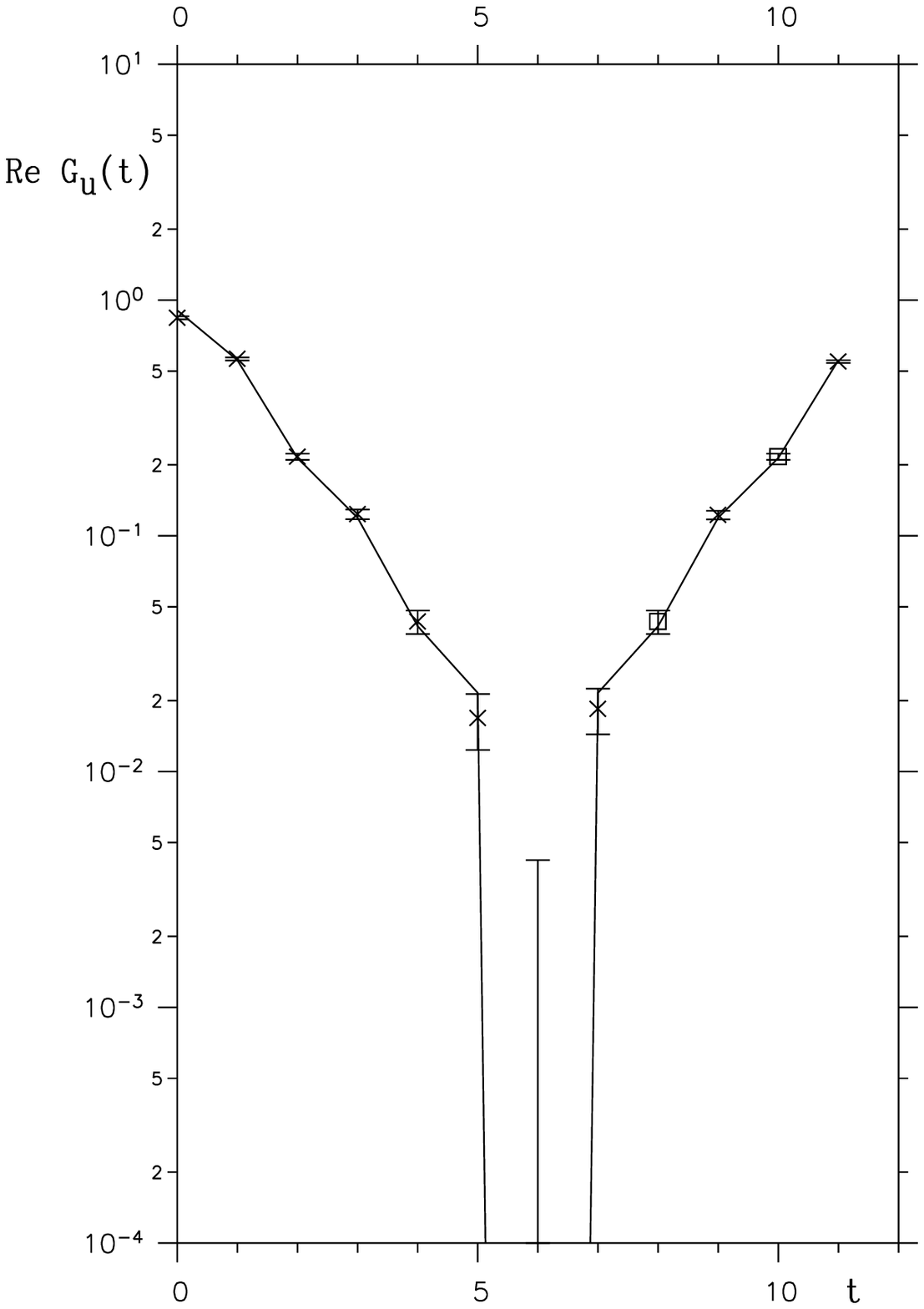}
\epsfysize=14cm
\epsfbox{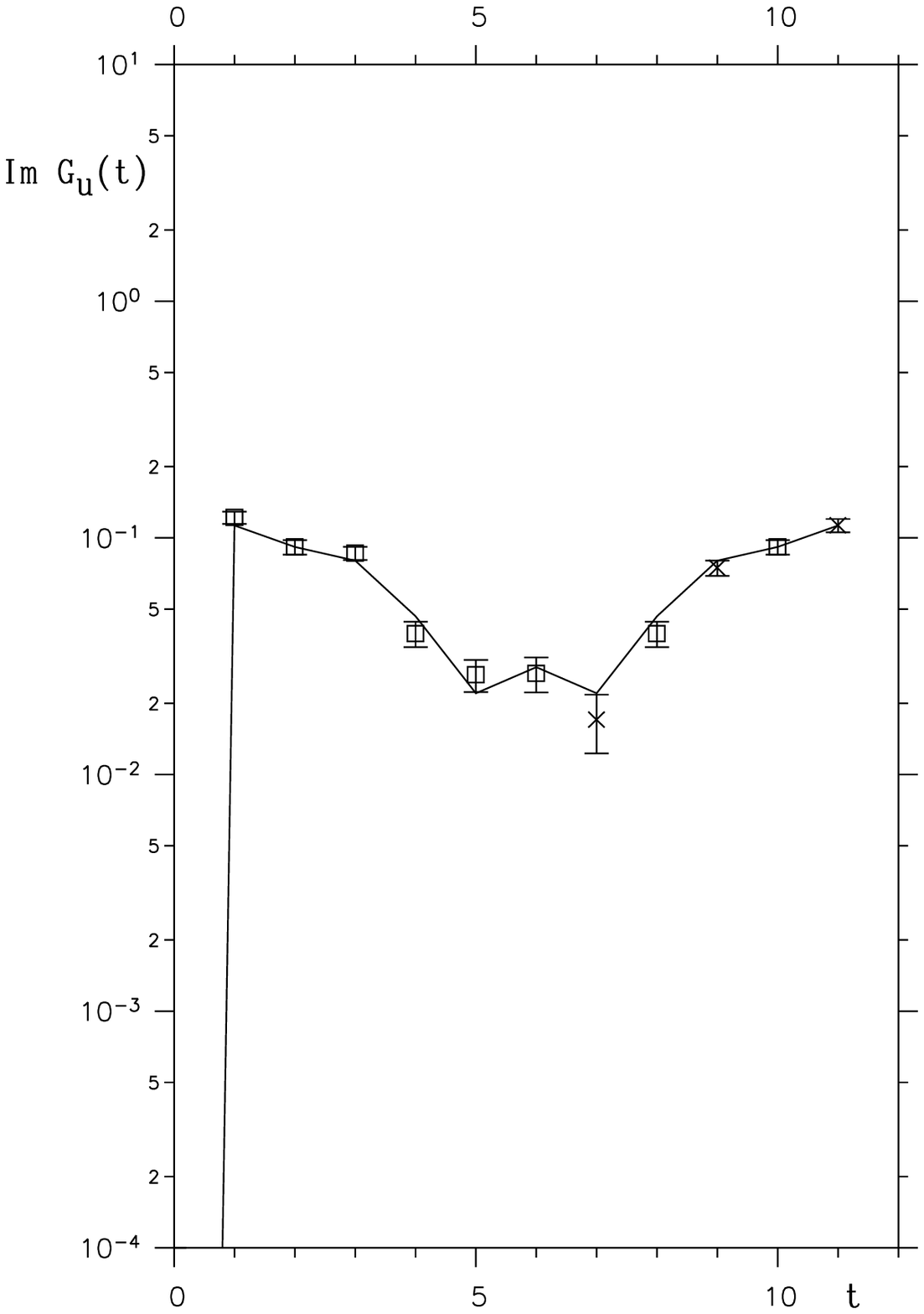}
}
\caption{$u$  fermion propagators at  $\beta = 0.16$
and $m_u = m_d =
  0.02$. Crosses correspond to positive values of the correlation function,
squares to negative values. Solid lines denote the fit.}
\label{fig:fermprop1}
\leavevmode
\vspace{-4.5cm}
\centerline{
\epsfysize=14cm
\epsfbox{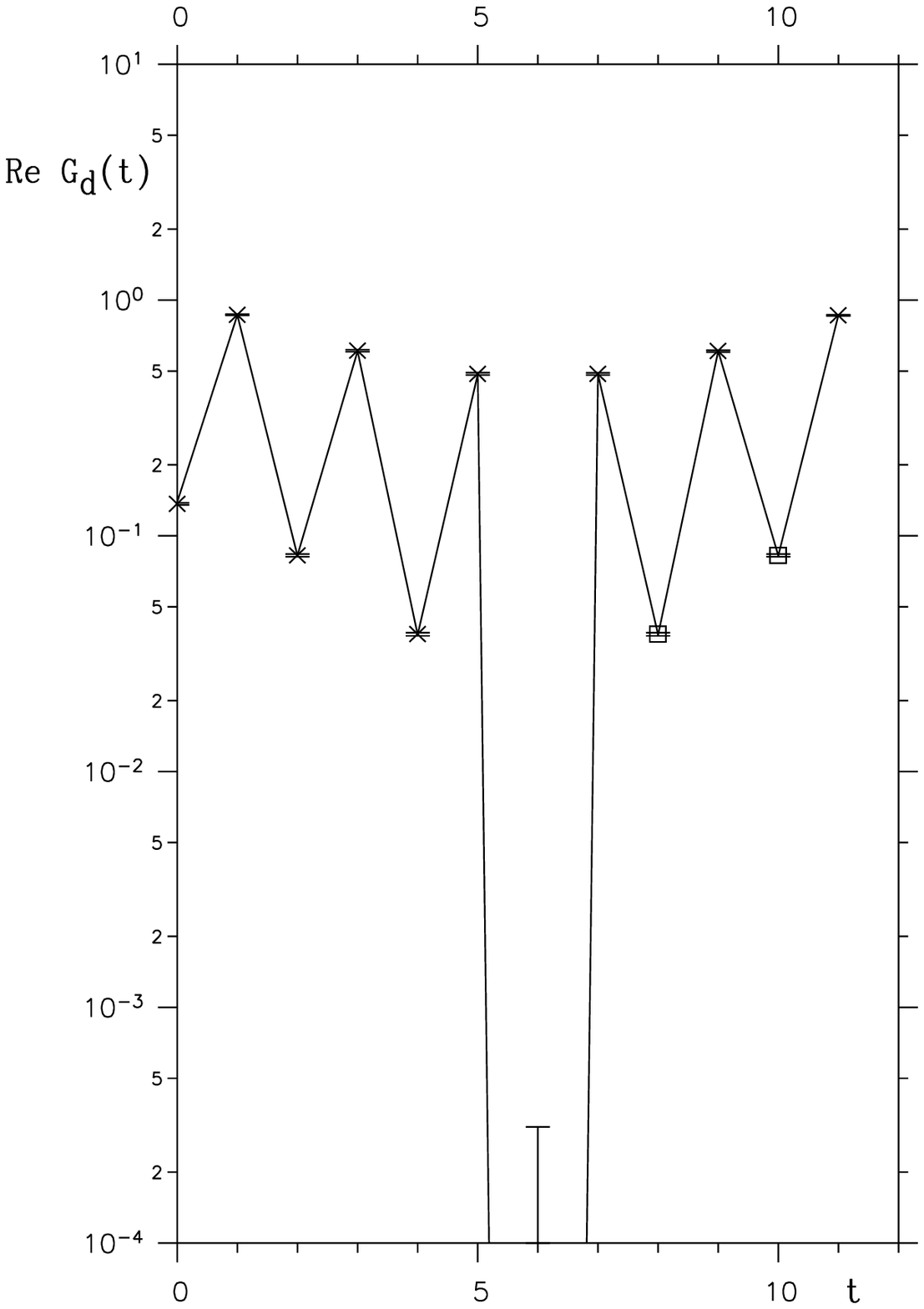}
\epsfysize=14cm
\epsfbox{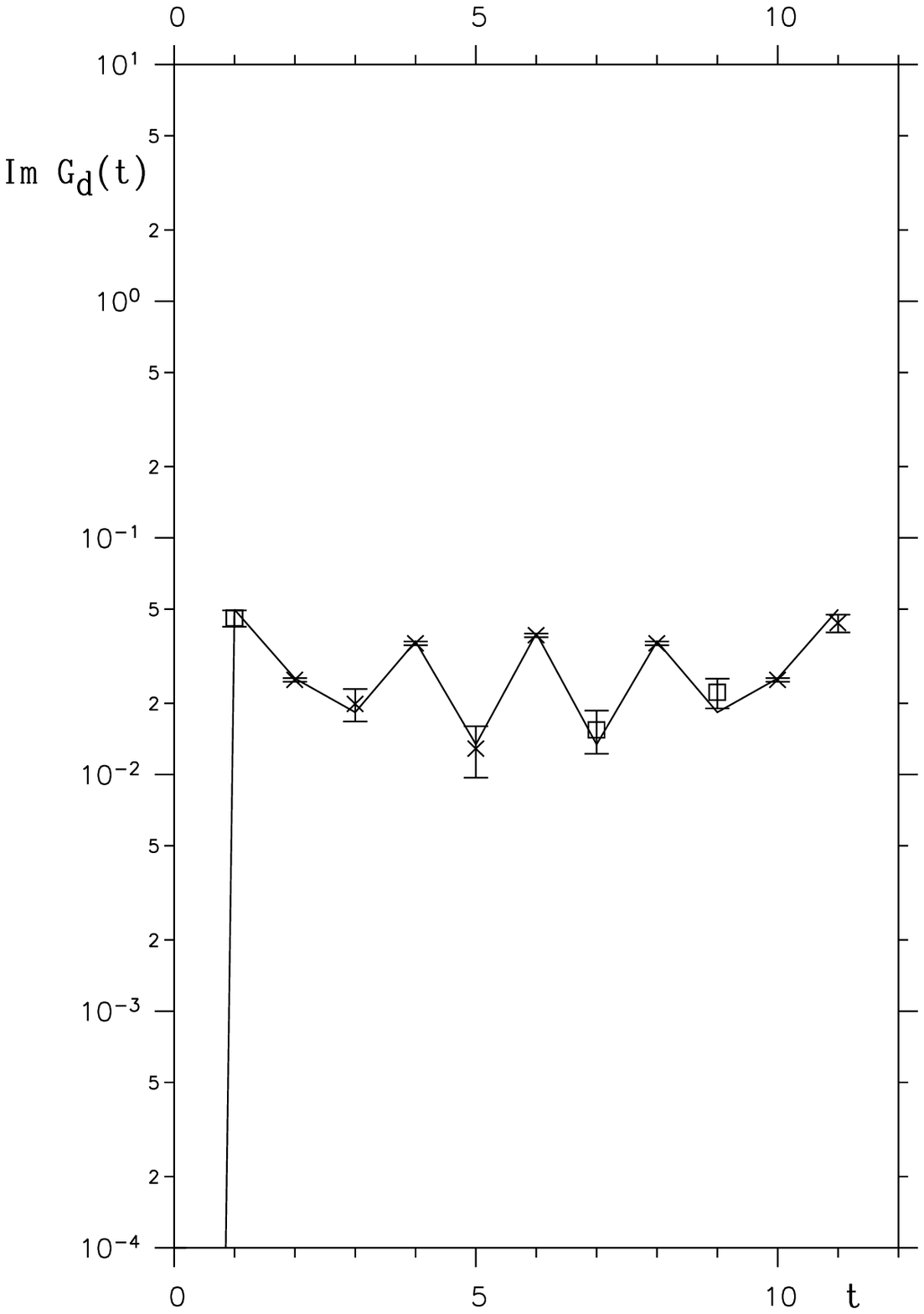}
}
\caption{ $d$  fermion propagators at  $\beta = 0.16$
and $m_u = m_d =
  0.02$. The meaning of the symbols is  explained above.}
\label{fig:fermprop}
 \end{figure}
The lattice average of the gauge field
\begin{equation}
\bar{A}_{\mu} = \frac{1}{V} \sum_{x} A_{\mu}(x), \; V = \prod_{\mu}L_{\mu},
\end{equation}
has a nonvanishing expectation value on our (relatively small) ensembles.
By shifting it by multiples of $4\pi/L_{\mu}$, such that it is restricted to
the interval $(-2\pi/L_{\mu},2\pi/L_{\mu}]$, this additional degree of
freedom was fixed. In the \twm\, this interval is  chosen
twice as large as in a
model with fermions of charge 1. This is necessary to preserve also the
boundary conditions of the $d$ fermion, which couples with charge $-1/2$.
$\bar{A}_{\mu}$ is approximately constant over $O(10-20)$
configurations. Following the procedure described
in~\cite{goeckeler91,schi92},
the set of data samples at each parameter value was divided into subsets of
10-20.  The  correlation functions were averaged over each subset
and fitted with the free form of a staggered fermion propagator in
a constant background field $B_{\mu}$. For the $u$ fermion,
$B_{\mu}$ agrees with the expectation value of
$\bar{A}_{\mu}$, taken over the given subset, for the $d$ fermion
with the expectation value of $-\bar{A}_{\mu}/2$. The fits were performed
with the routine MINUIT. The Ansatz gives a good
description of the data. An example for this is shown in
figure~\ref{fig:fermprop}. Results are given in
tables~\ref{tab_mr1} and~\ref{tab_mr2}.
Another indication that the fermion correlation functions behave
simliar to
free propagators with a renormalized mass $m_{u,d R}$, is obtained by
comparing the simulation data for the chiral condensates with
free propagator expressions:
\begin{equation}
 \langle \bar{\chi}_{k}\chi_{k}\rangle  =
 \frac{1}{V}\sum_{p_{\mu}}\frac{m_{kR}}{\sum_{\mu}\sin^{2}p_{\mu}
   +m_{kR}^{2}}, \; k=u,d,   \label{eq:gapqed}
\end{equation}
with the lattice momenta
\begin{eqnarray}
  p_{i}&=&\frac{2\pi}{L_{i}}n_{i}, \;
  n_{i}=1,\ldots,L_{i},\;i=1,\ldots,3;\nonumber \\
p_{4}&=&\frac{\pi}{L_{4}}(2n_{4}-1), \;
  n_{4}=1,\ldots,L_{4},\;\nonumber
\end{eqnarray}
\begin{figure}[thp]
\vspace{-5cm}

\epsfysize=14cm
\centerline{\epsfbox{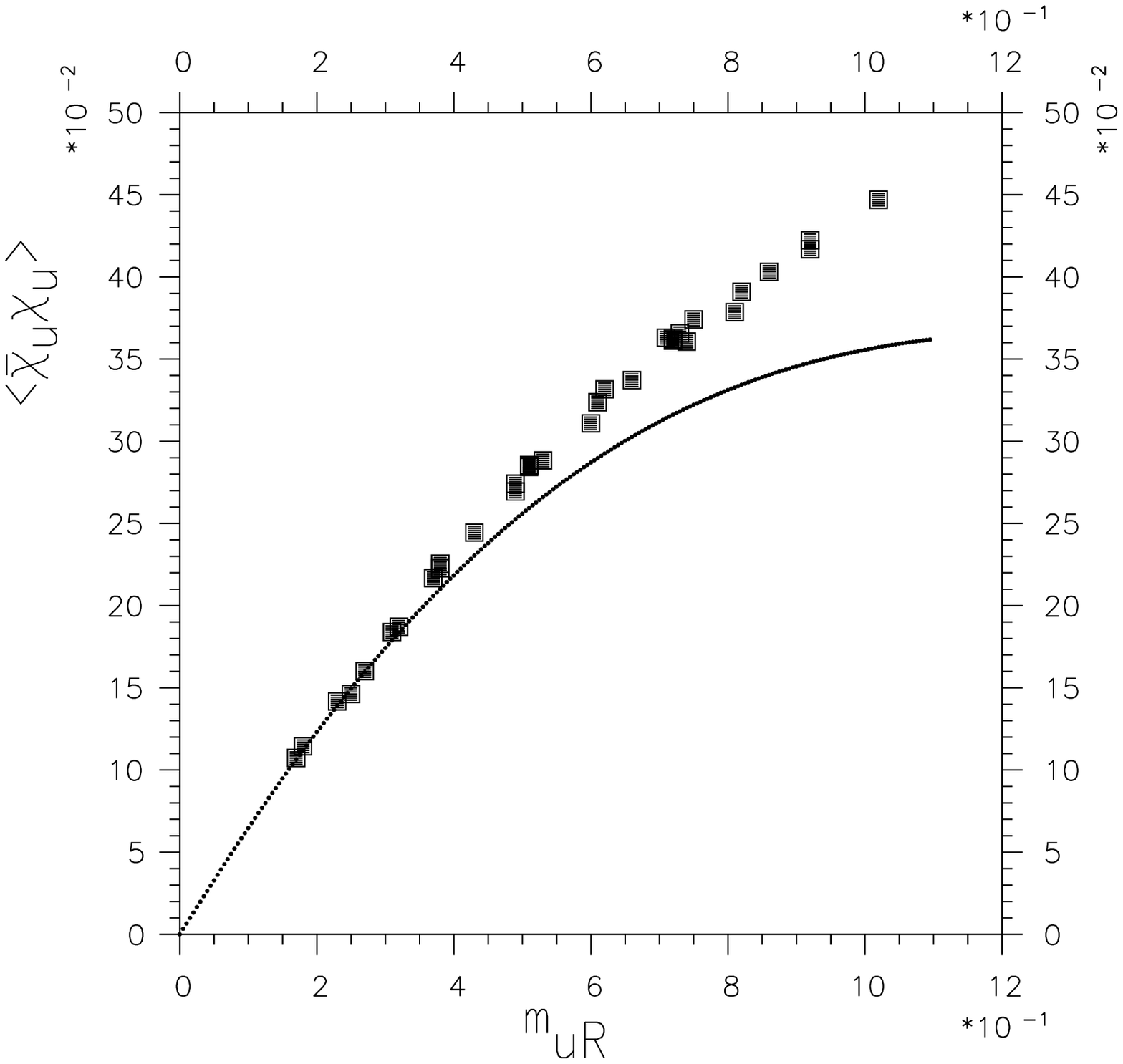}}
\vspace{-1.5cm}
\caption{ Chiral condensate $\langle \bar{\chi}_u\chi_u\rangle $ for $\beta
\geq 0.14$
plotted  against
the renormalized mass $m_{uR}$. The line denotes
eq.~(\protect\ref{eq:gapqed}).}
\label{fig:chimr01}
\vspace{-3cm}
\epsfysize=14cm
\centerline{\epsfbox{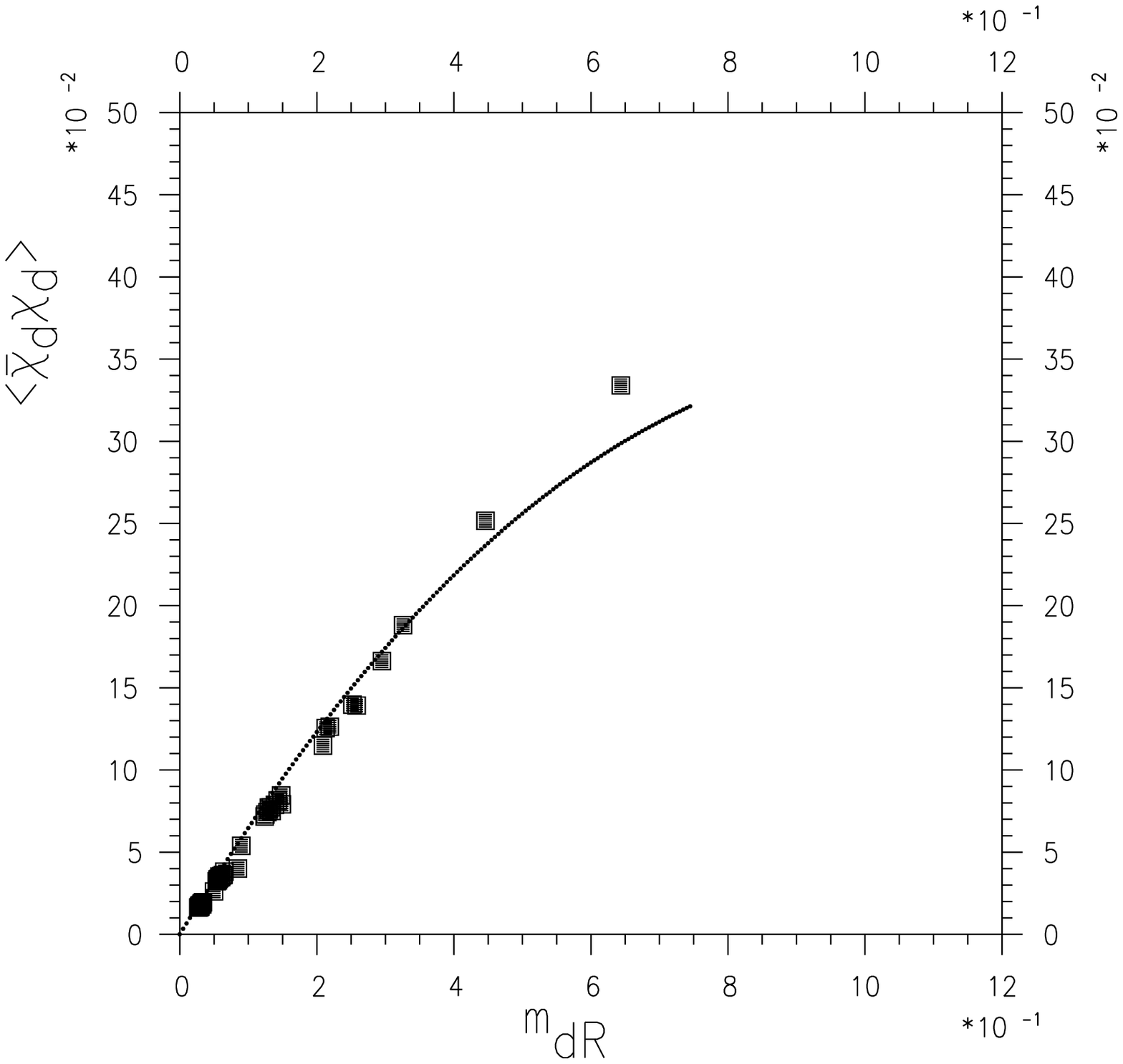}}
\vspace{-1.5cm}
  \caption{ Chiral condensate $\langle \bar{\chi}_d\chi_d\rangle $ plotted
against  $m_{dR}$.  The line denotes
eq.~(\protect\ref{eq:gapqed}). }
\label{fig:chimr02}
\end{figure}
Figures~\ref{fig:chimr01} and~\ref{fig:chimr02} show that for small masses
the results agree quite well with eqs. (\ref{eq:gapqed}).
 The fermion wave function renormalization constant is $O(1)$.
Since $m_{kR} \propto \langle \bar{\chi}_k\chi_k\rangle $ for small $m_{kR}$,
 eqs.~(\ref{eq:state}) and~(\ref{eq:cu}) imply that near  $\beta
\simeq \beta_{cu}$ the renormalized $u$ mass scales according to
\begin{equation}
m_{uR} \ln^{-1/3}(m_{uR}^{-1})\propto m_u^{1/3},\label{scal_mu}
\end{equation}
and the renormalized $d$ mass according to
\begin{equation}
 m_{dR} \ln^{-p_d}(m_{dR}^{-1})\propto m_d. \label{scal_md}
\end{equation}
The scaling behaviour of the $d$ is thus  in this region close to the
perturbative behaviour, which is very different from the behaviour of the $u$
fermion in this region.
For very large $\beta$ both renormalized masses follow eq. (\ref{scal_md}).
In the neighbourhood of $\beta \simeq \beta_{cd}$, eqs.~(\ref{eq:state})
and~(\ref{eq:cu}) indicate that
\begin{equation}
m_{dR} \ln^{-1/3}(m_{dR}^{-1})\propto m_d^{1/3}.
\end{equation}
In this region the difference between the bare and renormalized masses of
the $d$ becomes small.
\subsection{Phase Diagram}
\begin{figure}[bhtp]
\vspace{0.7cm}
\epsfysize=8cm
\centerline{\epsfbox{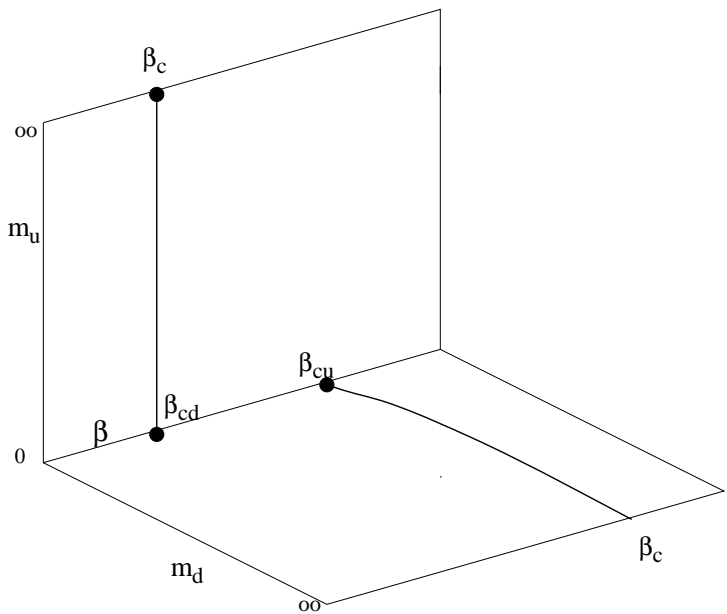}}
  \caption{Sketch of the phase diagram.}
 \label{fig:phase}
\vspace{-3cm}
\end{figure}

Figure~\ref{fig:phase} shows a sketch of the phase diagram. The agreement of
the results with eq. (\ref{eq:state}) suggests that there are two distinct
regions of chiral
symmetry breaking for both fermion species. This does not correspond to
the expectations in a confining theory.
If chiral symmetry breaking in non-compact QED was related to
confinement, one would expect both species of fermions to develop chiral
condensates in the same time.  The end
points of the phase boundary of the $u$,  which separates the
symmetric phase of the $u$  on the $m_u = 0$ surface from its broken phase,
are given by $\beta_{cu}$ and $\beta_c$, the critical point of QED with one
set of
$u$ fermions~\cite{schi92} which is the limit of the \twm\ if $m_d \rightarrow
\infty$.
In the presence of more charges, the critical point is shifted
towards stronger coupling, so $\beta_{cu}$ is slightly smaller than $\beta_c$.
 In
the limit $m_u \rightarrow \infty$, which corresponds to QED with one set of
charges with  coupling $\beta = 4/e^2$ one expects this to occur at
$\beta_c/4$ ($\sim 0.046$, using the result of~\cite{schi92}).
In the \twm\ a value very close to this is obtained, $\beta_{cd} \simeq
0.047$. Below $\beta_{cu}$,
no continuum limit with two fermion species  is possible.
 For the investigated parameter values at $\beta \leq 0.14$, the
renormalized $u$ masses are $O(1)$ in lattice units, which means the $u$
fermion is in this region
practically not present in the spectrum. In the region
$\beta \gtrsim \beta_{cu}$ both fermion masses can go to zero, so this
region is the most
interesting candidate for a continuum limit of the model.
\clearpage
\section{\label{sec:charge} The renormalized coupling}
\subsection{Charge determination on the lattice}
The Ward identities  ensure that the charge
renormalization is entirely determined by the wave function renormalization of
the photon:
\beqn
e_{u,d\,R}^{2} = Z_{3}\: e_{u,d}^{2}.
\eeqn
$Z_3$ is given by the zero momentum limit of the gauge invariant part of the
photon propagator:
\beqa
D(k)&=&\frac{\beta}{N_{k}V}\sum_{\stackrel{\scriptstyle \mu, k|k^2 = {\rm
      const},}{{k_{\mu} =
      0}}}
\hat{k}^{2}\left.\langle\tilde{A}_{\mu}(\hat{k})\tilde{A}_{\mu}(-\hat{k})
\rangle\right|_{k_{\mu}=0}, \label{eq:aa} \\
Z_{3}& =& \lim_{k \rightarrow 0} D(k) .
\eeqa
The sum in eq. (\ref{eq:aa}) runs over all directions $\mu$ and for each $\mu$
over all $N_k$ choices of $k$ with  fixed $k^2$ and $k_{\mu}=0$. Due to the
lattice, the momenta enter the photon propagator as
\beqn
\hat{k}_{\mu}=e^{ik_{\mu}}-1 \;\mbox{and}\;\hat{k}^{2} =
\hat{k}_{\mu}{\hat{k}}^{\ast}_{\mu}.
\eeqn
The right hand side of
(\ref{eq:aa}) turns out to be strongly fluctuating and inappropriate for an
extrapolation to $k \rightarrow 0$. Thus for the calculation of $D(k)$ a
method analogous  to references~\cite{schi90,schi92,horsley91} is used.
The photon propagator $D(k)$ is re-expressed
using the Ward identities~\cite{lusch90} in terms of a correlator between the
gauge field and the fermion current:
\beqn
D(k) = 1 - \left.\frac{1}{N_{k}V} \sum \left\langle\tilde{j_{\mu}}(\hat{k})
\tilde{A_{\mu}}(-\hat{k})\right\rangle \right|_{k_{\mu}=0},
\label{eqn:eichstrom}
\eeqn
where in the \twm\
\beqn
j_{\mu}(x) = \frac{\delta}{\delta A_{\mu}(x)}\sum_{yz}\left \{
(\bar{\chi}_{u}(y)M_{u,yz}\chi_{u}(z)+  \bar{\chi}_{d}(y)M_{d,yz}
\chi_{d}(z)\right\}. \label{eqn:curr}
\eeqn
The correlator in eq.~(\ref{eqn:eichstrom}) has less fluctuations and could
be used  for an extrapolation of $D(k)$ to $k \rightarrow 0$.
The fermion current was computed using a stochastic
estimator with 30-75 inversions of the fermion matrices.
\subsection{Comparison with perturbation theory}


Taking contributions of both fermions into account, in one-loop
perturbation theory the vacuum polarization tensor has the following form:
\begin{equation}
 1/\beta \; \Pi_{\mu\nu}(k,m_{uR},m_{dR},V) = e_u^2\;
\Pi^{(1)}_{\mu\nu}(k,m_{uR},V) + e_d^2\;
 \Pi^{(1)}_{\mu\nu}(k,m_{dR},V), \label{eqn:pmn}
\end{equation}
where $\Pi^{(1)}_{\mu\nu}(k,m_R,V)$ is the vacuum polarization tensor for a
single set of fermions with the renormalized mass $m_R$ on a lattice of
volume $V$. Projection onto the gauge invariant part yields
\begin{eqnarray}
  D(k) &=& 1 + e_u^2 \,\Pi^{(1)}(k,m_{uR},V) + e_d^2 \,\Pi^{(1)}(k,m_{dR},V),
\end{eqnarray}
where $\Pi^{(1)}(k,m_{R},V)$ is the one loop vacuum polarization function for
one set of staggered fermions:
\begin{equation}
       \Pi^{(1)} (k,m_R,V)
     =                    {1\over{\hat{k}^2}}
                          \left.\left[ \Pi_{\mu\mu}^{(1)}(k,m_R,V)
                                 - \Pi_{\mu\mu}^{(1)}(0,m_R,V)
                          \right]\right|_{k_{\mu} = 0}.
   \label{z3.m}
\end{equation}
The second term on the right hand side occurs because
$\Pi_{\mu\mu}^{(1)}(0,m_R,V) \neq 0$ for a finite $V$. This would correspond
to a finite photon mass, and  the term is subtracted off.
Here, a fixed $\mu$ has been chosen with $k_{\mu}=0$. Finally one obtains for
the photon propagator in the \twm:
\beqn
\frac{\beta}{D(k)} = \beta_{R} + \Pi(0,m_{uR},m_{dR},\infty)
-\Pi(k,m_{uR},m_{dR},V).
\eeqn
Extrapolating this to $V \rightarrow
\infty, k \rightarrow 0$:
\begin{equation}
  \beta_R = \beta - \Pi(0,m_{uR},m_{dR},\infty),\label{eq:vinfty}
\end{equation}
gives the perturbative relation between the bare and the renormalized coupling.
Combining the last two expressions, one gets the fit formula for the \MC\
results for $D(k)$:
\beqn
\frac{\beta}{D(k)} = \beta_{R} + \Pi(0,m_{uR},m_{dR},\infty)
-\Pi(k,m_{uR},m_{dR},V).
\eeqn
The renormalized coupling $\beta_R$ is the only free parameter in the fit.
\begin{figure}[htb]
\leavevmode
\vspace{-4cm}
\centerline{
\epsfysize=15cm
\epsfbox{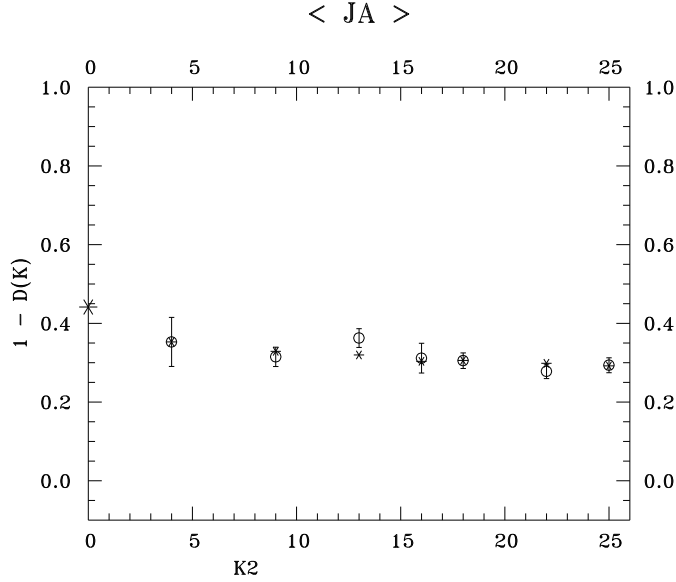}
}
\vspace{-2cm}
\caption{Photon propagator, $\beta$ = 0.17, $m_u$ = $m_d$ = 0.02.
The circles describe the data and the asterisks the fit with the
perturbative formula. The extrapolation to $k \rightarrow 0$ is
denoted by an asterisk. The variable $K2$ on the x axis  is defined as
$K2 = 576/(2\pi)^2k^2$ to be an integer.}
\label{fig:dk}
 \end{figure}

For the calculation of $\Pi(0,m_{uR},m_{dR},\infty)$, the effect of the zero
modes  $\bar{A}_{\mu}$ of  the gauge field has to be taken into account. Their
contribution is important where fermions are
light, as are the $d$ fermions at the simulated parameter values with
$\beta \sim \beta_{cu}$.
To calculate the perturbative vacuum polarization functions,
 the background fields are integrated over:
\begin{equation}
  \Pi(k,m_{uR},m_{dR},V) = \frac{\int d^4\bar{A} \det(D_u + m_u)\det(D_d + m_d)
  \sum_p \rho(p,k,\bar{A},m_{uR},m_{dR},V)}
{\int d^4 \bar{A} \det(D_u + m_u)\det(D_d + m_d)}.
\end{equation}
Also in the presence of constant background  fields the fermion determinant of
free fermions can
be written as a product of contributions with definite momentum:
\begin{equation}
 \det M_i =  \det(D_i + m_i) = \prod_{k_i}\left[m_i^2 +
 \sum_{\mu}\sin^2(k_{i,\mu}+c_i \bar{A}_{\mu})\right] ^{\frac{1}{2}}\!,
\; i = u,d,
\end{equation}
with $c_u$ = 1 and $c_d = -1/2$. Choosing $\mu  = 3$ in eq. (\ref{z3.m}) the
function $\rho$ looks as follows:
\begin{eqnarray}
  \rho &=& \sum_{i = u,d}\left[-2\frac{\sum_{\mu \neq
      3}\left(\cos(\tilde{k}_{i,\mu})\sin(p_{\mu}/2)\right)^2}
  {D(\tilde{k}_{i}-p,m_{iR})
    D(\tilde{k}_{i}+p,m_{iR})}\right. \nonumber \\
 & & \!\!\left. \vphantom{\frac{1}{2}}+ {\textstyle
\sin^2\left((\tilde{k}_{i,3} -p_3)/2\right)} \!\left\{
  \frac{1}{D(\tilde{k}_i - p,m_{iR})} - \frac{1}{D(\tilde{k}_i +
    p,m_{iR})}\right\}^2 \right]\cos^2(\tilde{k}_{i,3}),
\end{eqnarray}
where
\begin{equation}
  D(k,m_{R}) = m_R^2 + \sum_{\mu}{\textstyle\sin^2(k_{\mu}/2)}
\end{equation}
and
\begin{equation}
  \tilde{k}_u = k_u +\bar{A} \;\mbox{and}\;\tilde{k}_d = k_d - \bar{A}/2.
\end{equation}
The integral over the background fields was evaluated using a \MC\ method,
representing the background fields through Gaussian random numbers and
calculating the weight of the fermion determinant in the presence of these
background fields.
Figure~\ref{fig:dk} shows an example for a fit of $D(k)$ with one-loop
 perturbation theory in the presence of background fields. The
oscillations of $D(k)$ are caused by the dependence of $k$ on the direction
due to the asymmetry of the lattice.  The renormalized couplings are listed in
tables~\ref{tab_mr1} and~\ref{tab_mr2}.

\begin{figure}[pt]
\leavevmode
\vspace{-3cm}
\centerline{
\epsfysize=12cm
\epsfbox{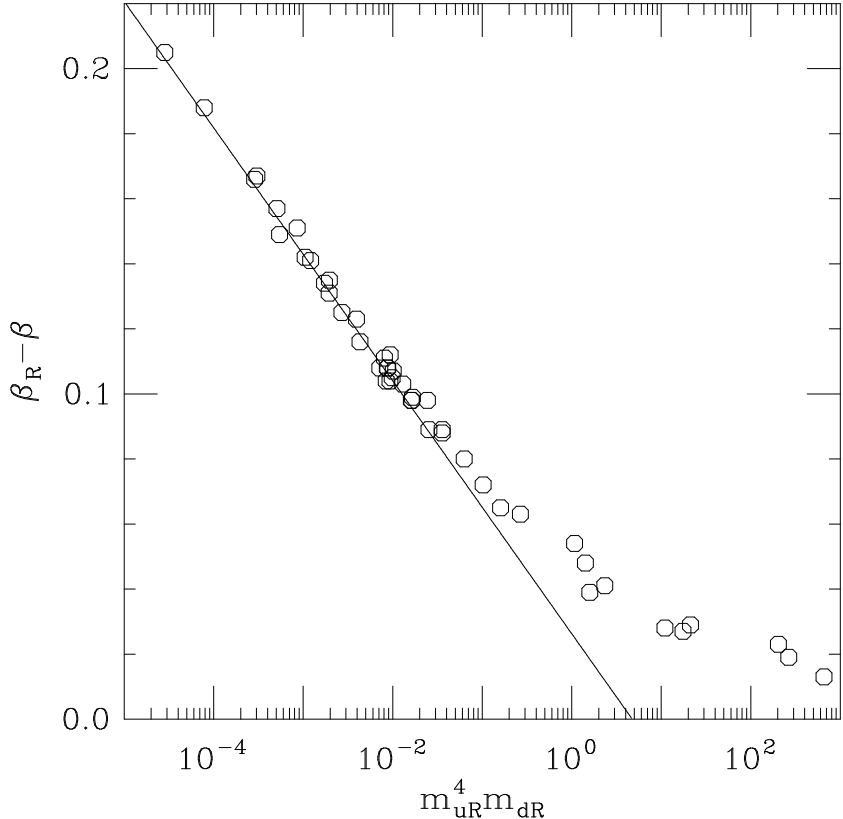}
}
\vspace{-1cm}
\caption{Renormalized couplings in the region of small $m_{uR}$ and
$m_{dR}$. The  line denotes the small mass limit of one-loop
perturbation theory.}
\label{fig:betaR}
\end{figure}

The renormalized coupling can be calculated directly in perturbation theory,
using eq. (\ref{eq:vinfty}). One would like to compare the data with the
perturbative result. Since this is of interest especially in regions of a
high cut-off, i.e. for small masses,
the limit $m_{uR}, m_{dR} \ll 1$ of the vacuum polarization
function~(\ref{eq:vinfty}) is taken, which gives
\begin{equation}
\beta_{R} - \beta = -\Pi(0,m_{uR},m_{dR},\infty) = -\frac{1 }{6 \pi^2}\ln
(m_{uR}^4m_{dR}) + 5c/4. \label{eq:sm}
\end{equation}
{}From QED with one species of fermions with charge 1, it is known that
$c\simeq 0.0210$~\cite{schi92}. Figure~\ref{fig:betaR} shows that in the small
mass limit the renormalized charges indeed follow the logarithmic behaviour as
in the right hand side of eq.~(\ref{eq:sm}). For larger masses, $\Pi$ is
no more a function of $m_{uR}^4m_{dR}$ alone, but the simulation results are
still in agreement with  perturbation theory~\cite{diss}. There does not
seem to be an effect on charge
renormalization from possible charged bound states.
The agreement with perturbation theory indicates that this model is trivial in
the limit of infinite cut-off, and that all renormalized charges vanish if only
one fermion mass goes to zero. The renormalized coupling vanishes in those
parts of the phase diagram (see figure~\ref{fig:phase}) where $\sigma_u$ or
$\sigma_d$ equal zero.

One would now like to find a relation
between a finite renormalized coupling and a corresponding cut-off. The ratio
of the renormalized masses has to be kept constant while the cut-off is
varied. Re-expressing (\ref{eq:sm}) in
terms of $R = m_{uR}/m_{dR}$ and exponentiating, one obtains:
\beqn
m_{uR} = 1.365 \: R^{\frac{1}{5}}
\exp{\left(-\left[\frac{6\pi^{2}}{5}(\beta_{R}-\beta)\right]\right)}.
\label{eq:mlim1}
\eeqn
{}From this equation one will be able to estimate the maximal cut-off belonging
to
each renormalized charge. For this, one needs to know which $\beta$ coordinate
the point with maximal cut-off on a \RGF\  line with fixed $\beta_R$ and $R$
has.
This flow line will be the intersection between the surface with fixed $R$
and the one with fixed $\beta_R$ in the three dimensional parameter space.

\subsection{Renormalization group flow of the coupling}
\begin{figure}[pthb]
\vspace{-5cm}
\centerline{\epsfxsize=10cm
\epsfbox{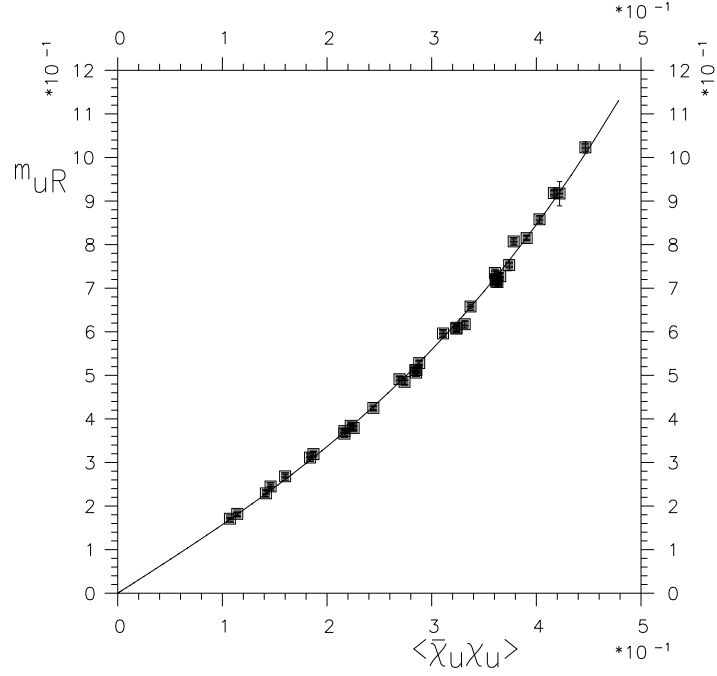}
}
\vspace{-4cm}
\centerline{\epsfxsize=10cm
\epsfbox{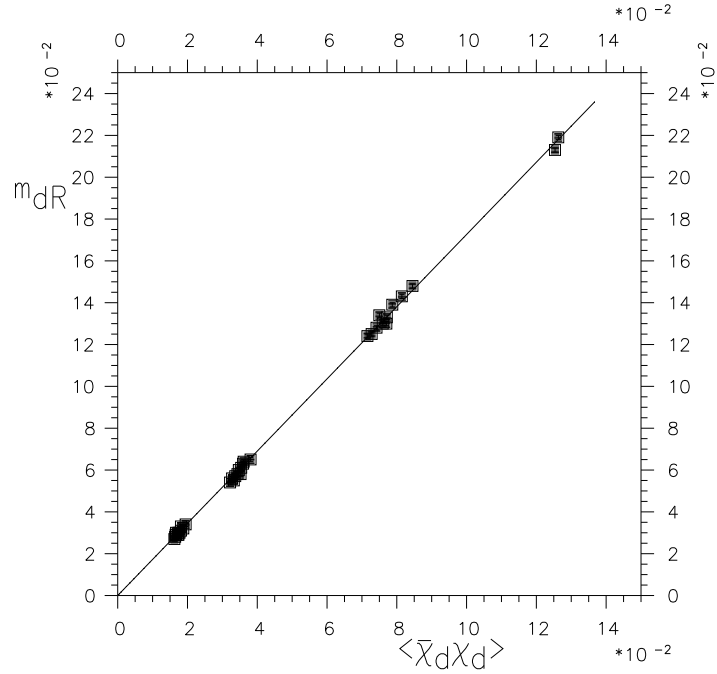}}
\vspace{-1cm}
  \caption{Renormalized masses as a function of the chiral condensates. The
squares denote simulation results and lines the fit according
to eqs.~(\protect\ref{eq:mrufit}) and~(\protect\ref{eq:mrdfit}),
respectively.}
 \label{fig:mrfit}
\end{figure}

It is not feasible to cover all the regions  of the
phase diagram that are of interest
with simulations. For a determination of the renormalization
group flows of the charge one therefore has to make use of extrapolation
methods. If one is able to find a functional dependence
of the renormalized mass on the bare parameters it is possible to
calculate the renormalized coupling using perturbation theory
from eq. (\ref{eq:vinfty}).
The chiral condensates can be approximated as functions of $\beta$, $m_u$ and
$m_d$
through the \eqss. Figures~\ref{fig:chimr01} and~\ref{fig:chimr02} suggest that
the renormalized masses can be expressed as functions of the chiral condensate,
where for small masses there is agreement with the free propagator relation
eq. (\ref{eq:gapqed}), and the leading term is linear.
For the renormalization group flows one is primarily interested in regions
where both fermion masses are smaller than 1, so $m_{uR}$ is fitted for
$\beta \geq 0.14$ with the Ansatz
\begin{eqnarray}
  m_{uR}& =& P_1\chiu    +  P_2\chiu^{3} . \label{eq:mrufit}
\end{eqnarray}
Here, $\chi^2$ per degree of freedom is 1.5. The following fit
gives a good description for the $d$ mass for $\beta \geq 0.14$:
\begin{eqnarray}
  m_{dR}& =& Q_1\chid, \label{eq:mrdfit}
\end{eqnarray}
with $\chi^2$ per degree of freedom equal to 2.2. A  cubic term is not needed
here. The fit parameters $P$
and $Q$ are given in table~\ref{tab:mrfit}. As illustrated in
figure~\ref{fig:mrfit}, one thus obtains a good description for
the renormalized masses.
To obtain the renormalized masses as a function of the bare parameters,
first the \eqss\ were solved explicitly for the chiral condensates using
M\"uller's
method~\cite{muller56} for calculating zeros of non-linear functions. The
results were thus inserted into the fit equations (\ref{eq:mrufit}) and
(\ref{eq:mrdfit}) respectively.

\begin{figure}[bthp]
\leavevmode
\vspace{-5cm}
\centerline{
\epsfysize=15cm
\epsfbox{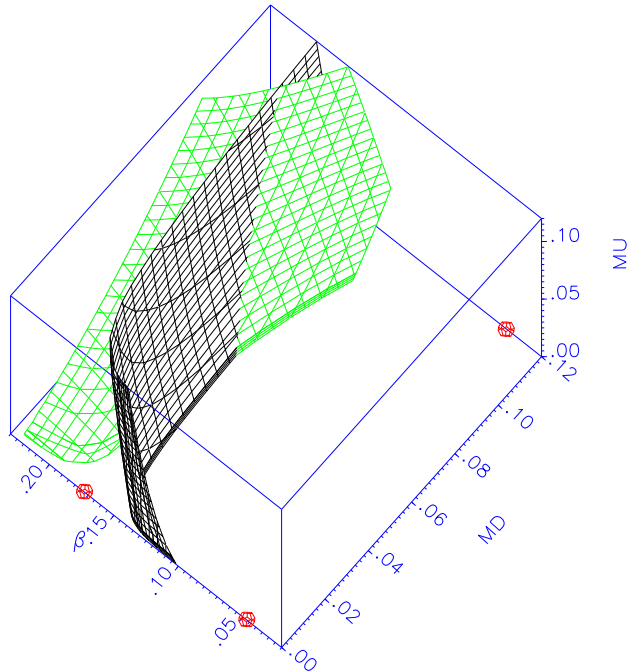}
}
\vspace{-1cm}
\caption{Surfaces with $\beta_R = 0.27$ (black) and $R = 5$
  (grey). The little spheres denote $\beta_{cu}$ and $\beta_{cd}$.}
\label{fig:flow1}
\end{figure}

Using this method, a grid of values for the mass
ratio $R$ in the phase diagram is generated and, using renormalized
perturbation theory, also for $\beta_R$. Using a graphics package,
surfaces of constant $\beta_R$ and $R$ can be drawn.
Figure~\ref{fig:flow1} shows a surface $S_{\beta_R}$ (black)
with a  constant $\beta_R$ and its intersection with a surface
$S_R$ (grey) of constant $R$ of $O(1)$.

For large $m_u$ and $m_d$  both renormalized fermion masses are large.
Therefore $\beta_R \sim \beta$, which implies that
surfaces with fixed $\beta_R$ are nearly perpendicular
 to the $\beta$ axis. The behaviour at small $m_u$, $m_d$ can be read off from
eq.~(\ref{eq:sm}). If $m_u$ is lowered, the surfaces bend over until they
end on the  $m_u = 0$
plane below the line where the chiral symmetry of the $u$ fermion is broken.
The larger $\beta_R$ is, the closer the end line of the surface on the $m_u=0$
plane comes to the critical line of the $u$, except for very small $m_d$.
In a whole range of $\beta <
\beta_{cu}$ and small bare masses, $m_{dR}$ is much smaller than $m_{uR}$ and
for constant $m_d$ nearly constant.  So surfaces with constant $\beta_R$
 bend at small  $m_d$ towards small $\beta$,
as the black surface in figure~\ref{fig:flow1} indicates. They intersect
the $m_d = 0$ plane in the broken phase
of the $d$ fermion, and thus cut the $\beta$ axis at
$\beta < \beta_{cd}$. The important point to note is that  $S_{\beta_R}$
intersects the surfaces  $m_u=0$ or $m_d=0$ only in regions where both
renormalized fermion masses are non-zero.

Surfaces of constant $R$ end for $\beta \geq \beta_{cu}$ on
the $\beta$ axis. The scaling behaviour of the masses in eqs.~(\ref{scal_mu})
 and (\ref{scal_md}) implies that at $\beta = \beta_{cu}$ these surfaces obey
the relation $m_u \sim c m_d^3$, where $c$ is a constant.
If $\beta$ is lowered past $\beta_{cu}$,
$m_{uR}$ increases quickly. As indicated in
figure~\ref{fig:flow1}, the end line of the surfaces with constant $R$ on the
$m_u = 0$ plane therefore has to turn away from the $\beta$ axis towards a
finite $m_d$. For an estimate of the validity bound, the intersection line of
both surfaces is of interest, as it represents a flow line where both $R$ and
$\beta_R$ are constant while the cut-off scale is changed.
 It ends on the $m_u = 0$ plane  at $\beta < \beta_{cu}$.
As one moves down along the intersection line, the cut-off becomes larger, but
never reaches infinity.

\subsection{Validity bound}
In non-compact QED with four flavours of staggered fermions of charge 1, the
$\beta$ function
goes, in the limit of small renormalized mass, over into the prediction of
one loop continuum perturbation theory for four fermion flavours of charge
1~\cite{schi92}.
In the \twm\, with four flavours of charge 1 and four
 flavours of charge $-1/2$,
 a corresponding relation is fulfilled~(\ref{eq:sm}). This
suggests that
in the presence of $N$  species of fermions the renormalized charge
can be approximated with  the formula
\beqn
\beta_R - \beta = -\sum_{i = 1}^N\frac{Q_i^2}{6\pi^2}\ln\left(am_{iR}\right)
+ \sum_{i=1}^N Q_i^2 c.
\eeqn
In this section the lattice spacing $a$ is kept explicitly, to illustrate the
dependence on the cut-off.
In QED with one charge and in the \twm\ it was reasonable to approximate the
bare charge corresponding to the point of maximal cut-off by the critical
coupling of the fermion with the strongest coupling. This critical coupling
is dependent on the number of dynamical fermions. Comparing $\beta_c$,
determined using mean field \eqss\ or mean field \eqss\ with logarithmic
corrections, in
the quenched case, in models with four flavours of charge 1~\cite{schi92},
with eight flavours of charge 1~\cite{kogut89}
and the \twm,  one finds the following behaviour:
\beqn
\beta_c(N\mbox{ species}) = \beta_c(\mbox{quenched}) - \epsilon\sum_{i=1}^N
 Q^2_i,
\eeqn
where $\epsilon \simeq 0.01
$.

In the Standard Model, one has three generations of fermions of charge 1,
three generations and  colours of charge $2/3$ and three generations and
 colours
of charge $1/3$. Expressing the electron mass in units of the cut-off and the
other fermion masses in terms of ratios with the electron mass, one obtains
\begin{equation}
 a m_{eR} \gtrsim 3.68 \exp\left\{-\frac{3\pi^2}{4}\beta_R\right\}\times
\left(R_{\mu}R_{\tau}\right)^{1/8}\left(R_u R_c R_t\right)^{1/6}\left(R_d R_s
R_b\right)^{1/24},
\end{equation}
where $R_{\mu}= m_{eR}/m_{\mu R}$ etc.
The relation between the lattice spacing $a$ and the
cut-off $\Lambda$  is approximately given by the relation:
\begin{equation}
  \Lambda \simeq 1/a.
\end{equation}
Using the physical fermion masses and the
physical
value of the fine structure constant $1/137$, one gets the cut-off
\begin{equation}
  \Lambda \lesssim  10^{32}\; \mbox{GeV}.
\end{equation}
This is much larger than the Planck scale  ($10^{19}$ GeV). However, if there
are more charged
particles, e.g. due to supersymmetry, the exponential dependence on the number
of charged particles might cause this cut-off to become considerably lower.

If on the other hand one sets the validity bound of QED to be at the Planck
scale one obtains the upper bound on the fine structure
constant
\begin{equation}
  \alpha_R \lesssim 1/80,\label{eq:bound}
\end{equation}
which is surprisingly small. One also has to note that in eq.
(\ref{eq:bound}) the effect of charged $W^{\pm}$ bosons is not yet included.
\section{\label{sec:comp} Composite states}
\subsection{Lattice operators and fits}
Correlation functions of scalar and pseudoscalar $\bar{u}u$ and $\bar{d}d$
states were investigated:
\begin{eqnarray}
 C_{u,i}(t) =  \left\langle\sum_{\vec{x}}s_{i}(\vec{x},t)\bar{\chi}_u(\vec{x},
t)
  \chi_u(\vec{x},t)\bar{\chi}_u(0)\chi_u(0)\right\rangle ,\;\;i = 1,\;2, \\
C_{d,i}(t) =  \left\langle\sum_{\vec{x}}s_{i}(\vec{x},t)\bar{\chi}_d(\vec{x},t)
  \chi_d(\vec{x},t)\bar{\chi}_d(0)\chi_d(0)\right\rangle,\;\;i = 1,\;2 .
\end{eqnarray}
The sign factors $s_{i}$ determine the lattice representation a state
belongs to~\cite{golterman86}. The corresponding continuum quantum numbers and
states in QCD terminology are listed in table~\ref{tab:localmes}.
To reduce statistical fluctuations, for each correlation function 32
local sources distributed over the lattice were used.
Masses were determined by fits to the formula:
\begin{equation}
 C(t)  = A_1 \left(e^{-E_1 t} + e^{-E_1(T - t)} \right)
   + A_2 (-1)^t \left(e^{-E_2 t} + e^{-E_2(T - t)} \right)
   + A_3 + (-1)^t A_4.
   \label{eq:fitnemesqed}
\end{equation}
Here, $E_1$ is the energy of the lightest pseudoscalar and $E_2$
the energy of the lightest scalar state. The constants $A_3$ and $A_4$
correspond to single fermions which propagate  in the time
direction around the lattice (see~\cite{NJL}). Here they were
not needed for fits of
$\bar{u}u$ states in any parameter region investigated.  Masses of neutral
composite states for $\beta \sim \beta_{cu}$ are presented in
table~\ref{tab_pi1} and for $\beta \sim \beta_{cd}$ in table~\ref{tab_pi2}.

\begin{figure}[p]
\leavevmode
\vspace{-5.5cm}
\centerline{
\epsfysize=14cm
\epsfbox{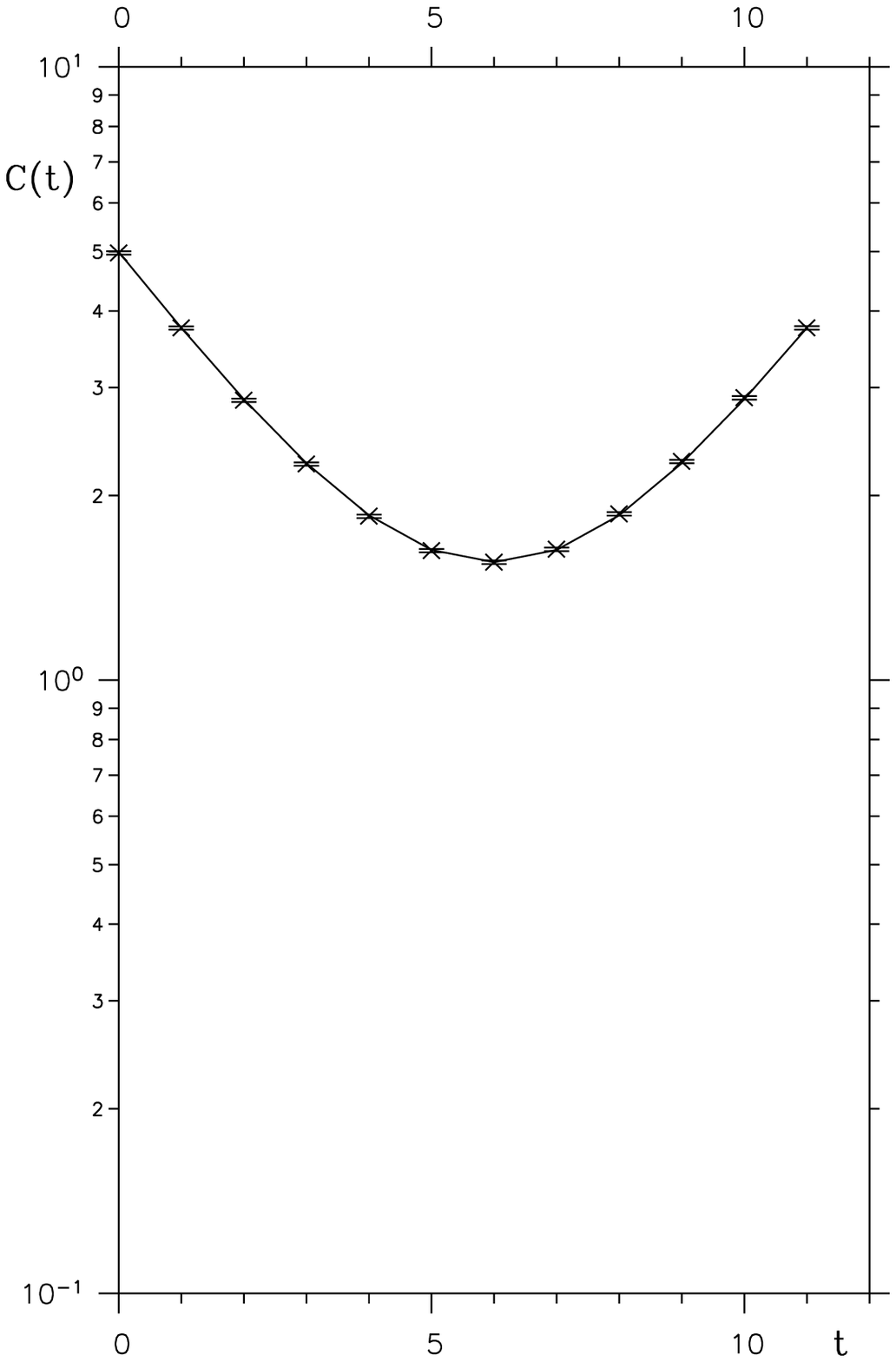}
\epsfysize=14cm
\epsfbox{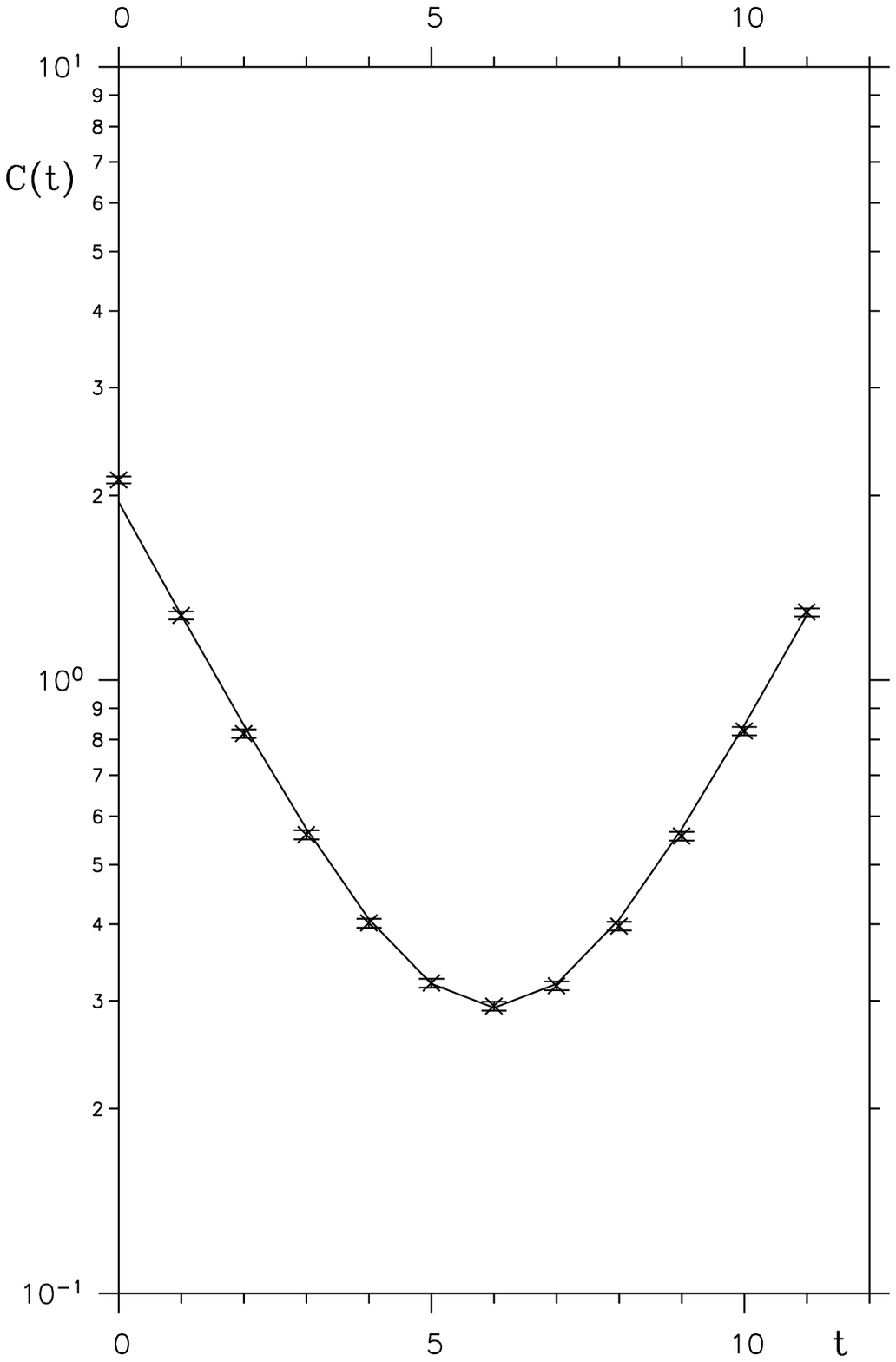}
}
\caption{Correlation functions type 2, $\beta$ = 0.05, $m_u$ = $m_d$ = 0.02,
  $\pi_u$ (left) and $\pi_d$ (right).
 Crosses denote positive
values of the correlation function, the solid line denotes
the fit.}
\label{cor1}
\leavevmode
\vspace{-4.5cm}
\centerline{
\epsfysize=14cm
\epsfbox{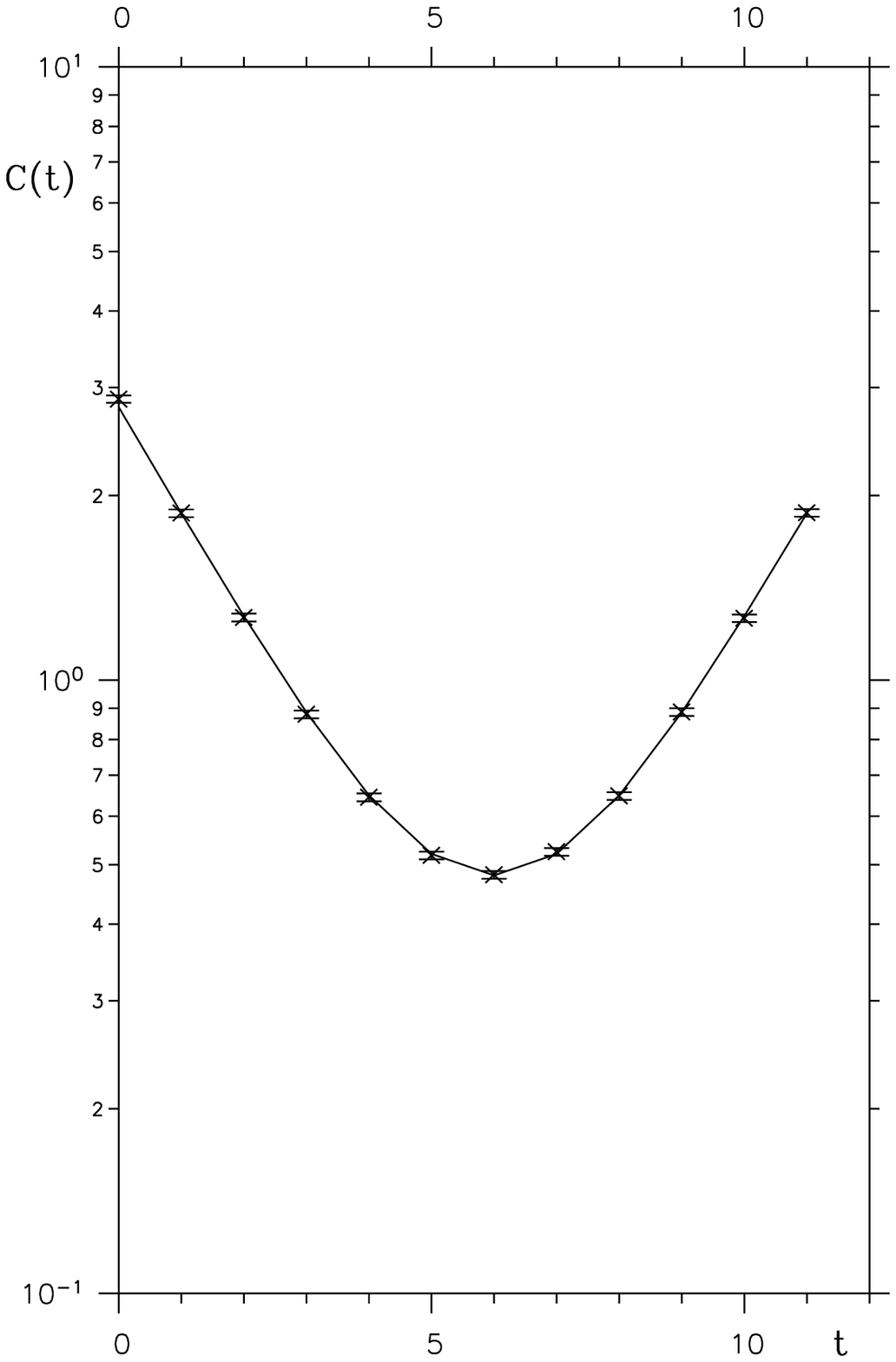}
\epsfysize=14cm
\epsfbox{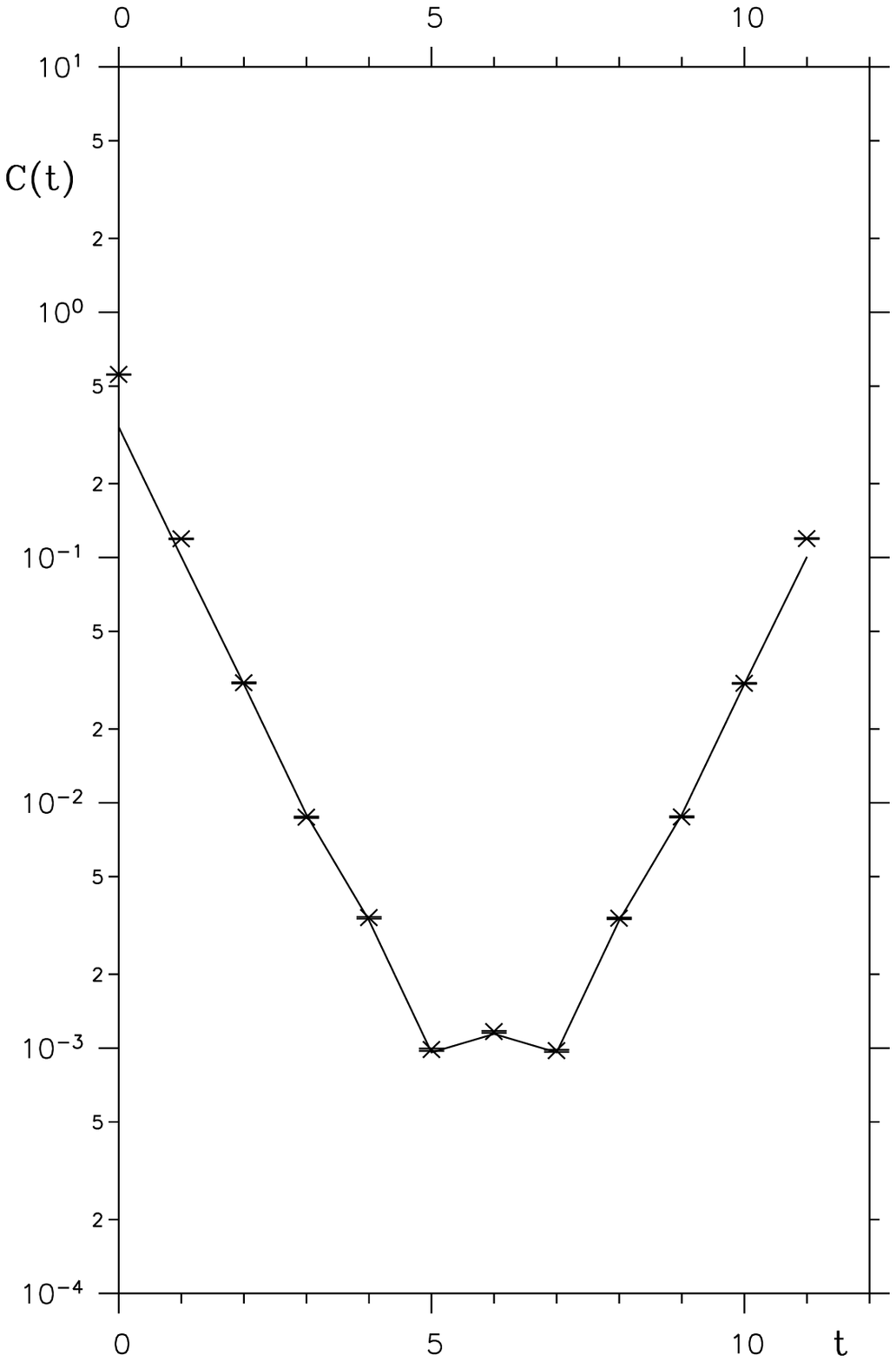}
}
\caption{Correlation functions type 2, $\beta$ = 0.17, $m_u$ = $m_d$ = 0.02,
  $\pi_u$ (left) and $\pi_d$ (right). The symbols have the same meaning as
above.}
\label{cor2}
 \end{figure}

For correlation  functions of type 2, fits were
done with $A_2$ set to zero. In this channel only the Goldstone pion
contributes, since states with the quantum numbers $0^{+-}$ cannot be
realized in the quark model.  Figure~\ref{cor1} shows correlation
functions of type 2 at $\beta \sim \beta_{cd}$ and figure~\ref{cor2} at
$\beta \sim \beta_{cu}$. For
$\bar{u}u$ states a fit interval $t_{min}/t_{max}$ = $1/11$ was chosen.
Close to $\beta_{cd}$ the mass of the $\pi_u$ is nearly
independent of the fit interval, which indicates that there is a good
overlap of the pointlike interpolating field with the pion.
For $\beta \leq 0.19$, the masses of the $\bar{u}u$ pions are smaller than
$2m_{uR}$, thus one can speak of bound states, which become lighter the deeper
one goes into the broken phase.  For $\beta \geq 0.20$ the $\pi_u$ masses lie
above $2m_{uR}$. This is an indication that the pion is not bound any more,
instead there is possibly in the infinite volume limit a resonance
in the pseudoscalar channel. It has to
be noted that when a spectrum with many states is fitted with a single
exponential,
the fit result might be an average between the lowest lying state around
$2m_R$ and the excited states~\cite{NJL}.

For $\bar{d}d$ states  the fit range was at
$\beta < 0.10$ chosen to be $t_{min}/t_{max}$ = $1/11$. For larger $\beta$,
good fits could not
be  obtained unless the constants $A_3$ and $A_4$ were included. $A_3$ and
$A_4$ are about $O(10^{-3})$ smaller than $A_1$. The fit interval was in this
$\beta$ range chosen to be $t_{min}/t_{max}$ = $2/10$.
The $\pi_d$ masses are around $\beta_{cd}$  below $2m_{dR}$, for
$\beta > 0.05$  there is no bound pseudoscalar $\bar{d}d$ state.

\begin{figure}[p]
\leavevmode
\vspace{-5.5cm}
\centerline{
\epsfysize=14cm
\epsfbox{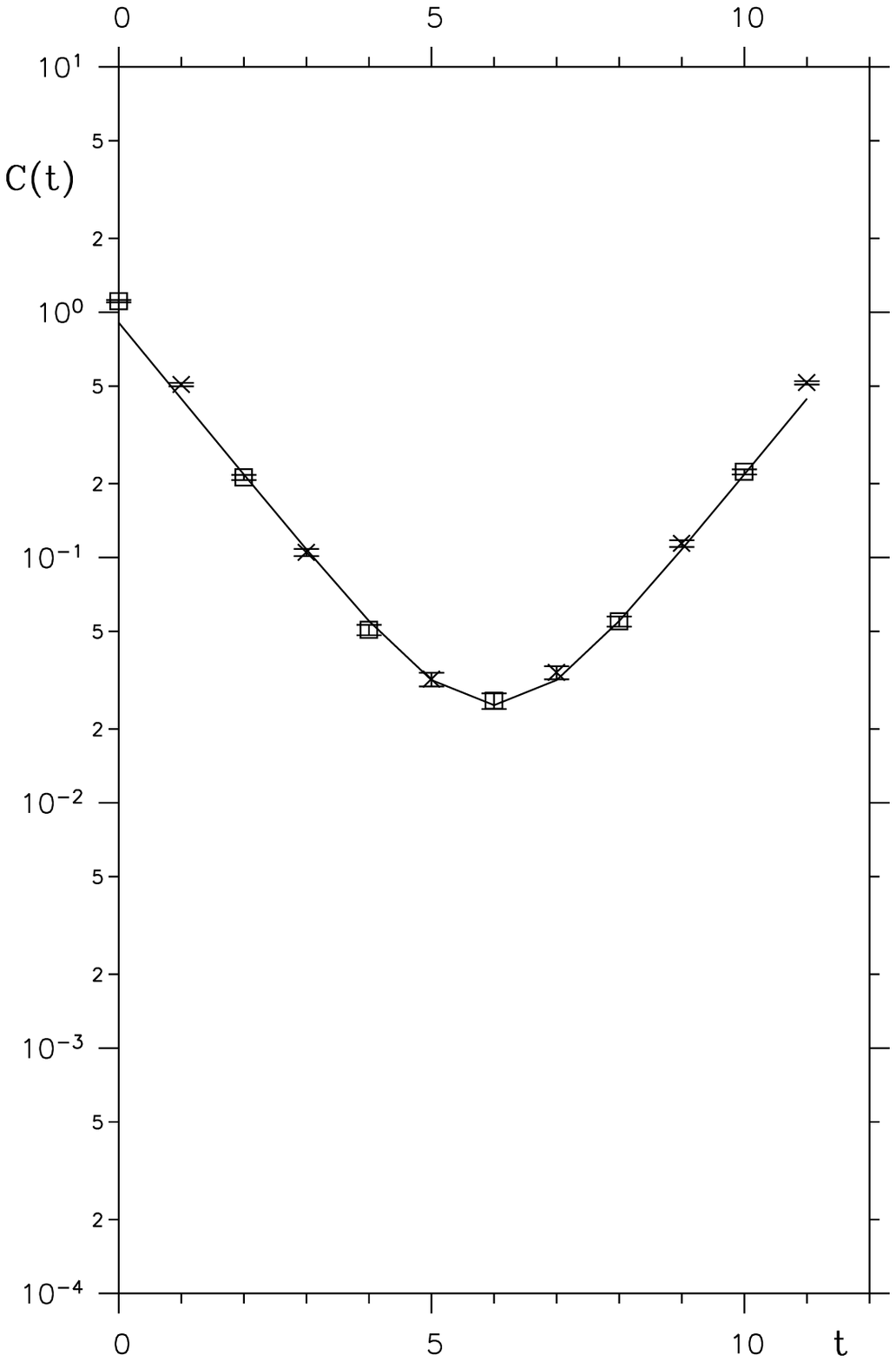}
}
\caption{Correlation functions type 1, $\beta$ = 0.05, $m_u$ = $m_d$ = 0.02,
  $\hat{\sigma}_d$.  Crosses denote positive and
squares negative values of the correlation function, the solid line denotes
the fit.}
\label{cor3}
\leavevmode
\vspace{-4.5cm}
\centerline{
\epsfysize=14cm
\epsfbox{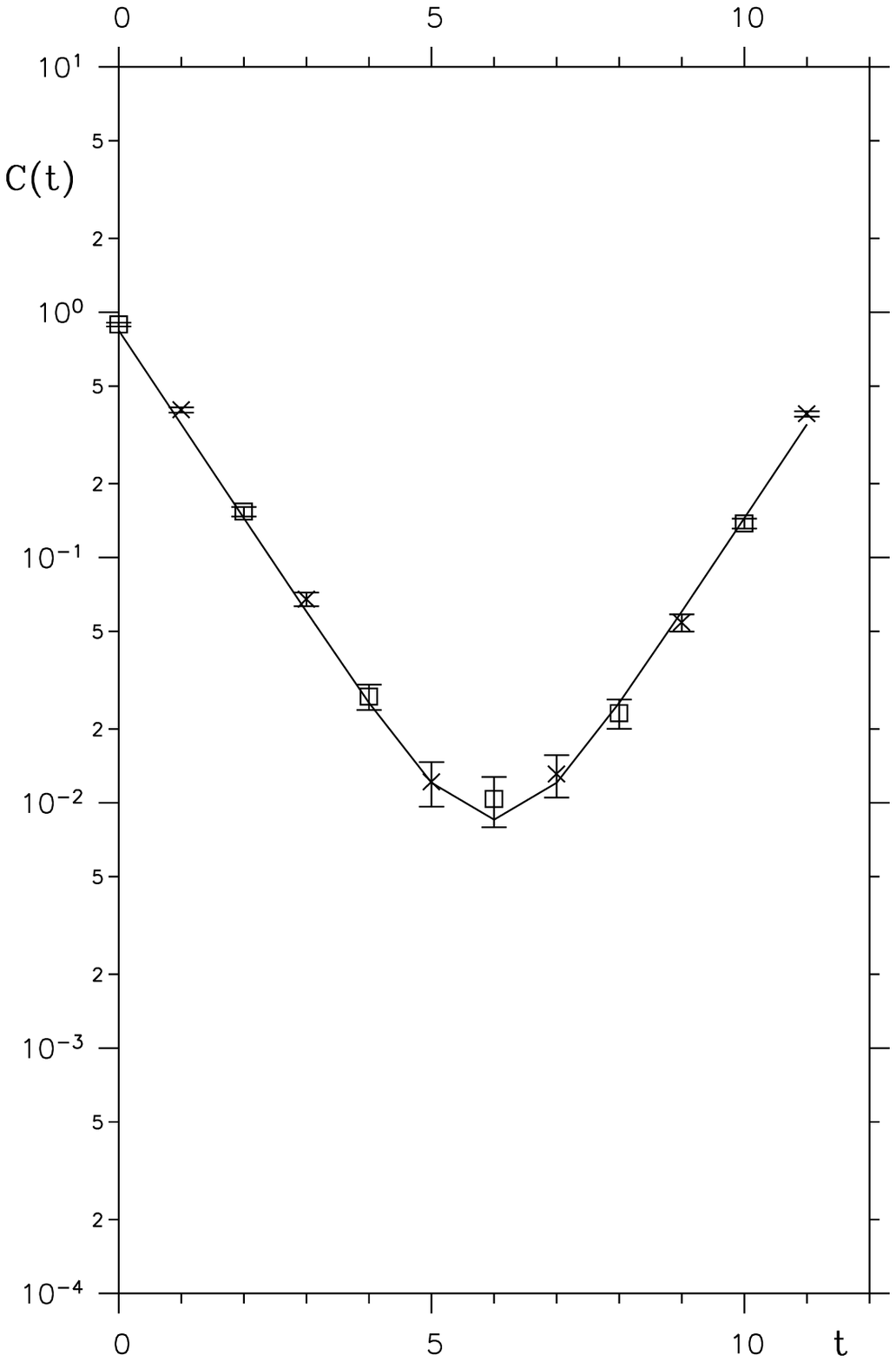}
\epsfysize=14cm
\epsfbox{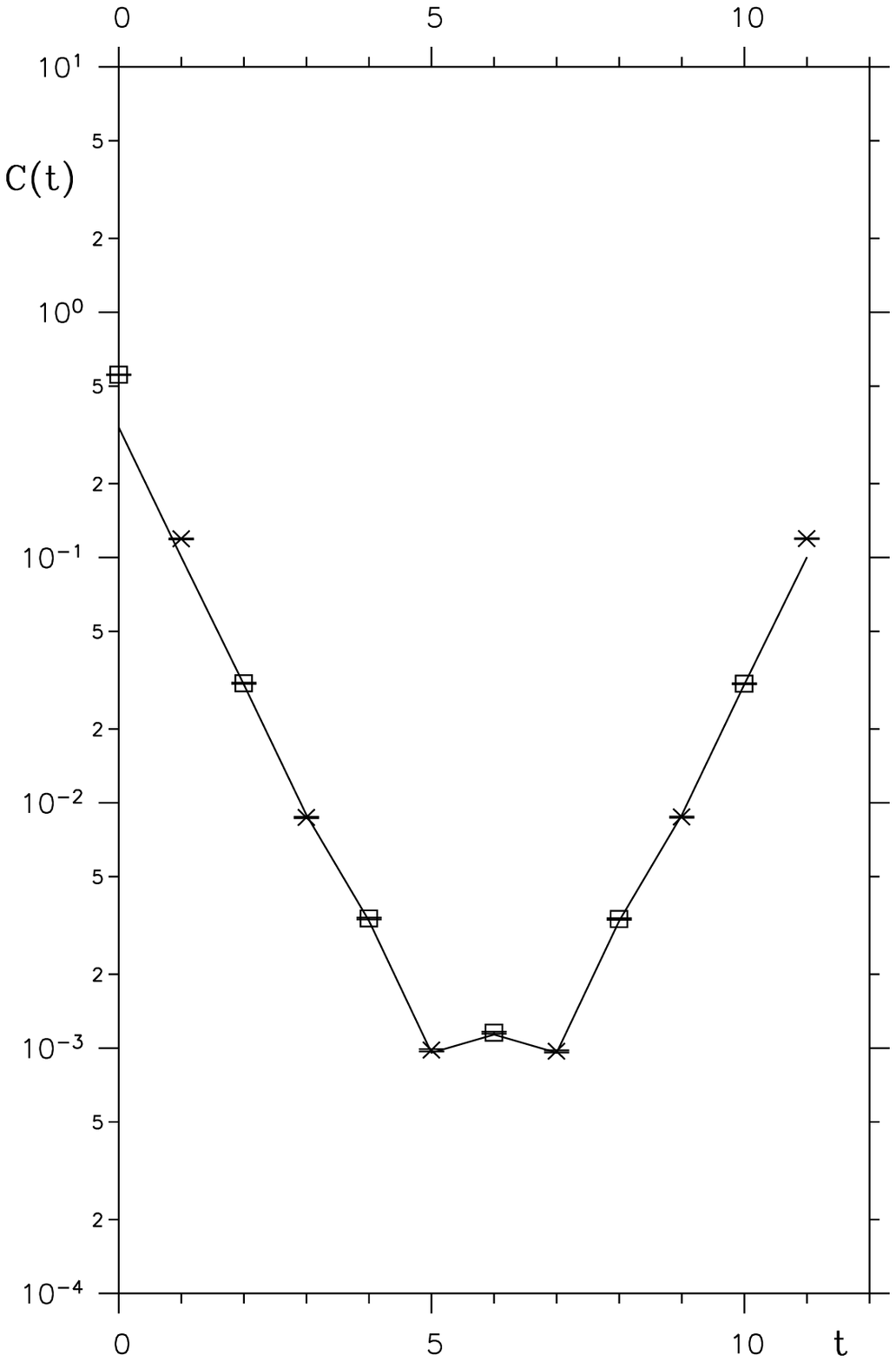}
}
\caption{Correlation functions type 1, $\beta$ = 0.17, $m_u$ = $m_d$ = 0.02,
  $\hat{\sigma}_u$ (left) and $\hat{\sigma}_d$ (right). The  symbols have
the same meaning as above.}
\label{cor4}
 \end{figure}

Figures~\ref{cor3} and~\ref{cor4} show typical correlation functions of type 1
for
couplings close to $\beta_{cu}$ and $\beta_{cd}$.  For $\beta$ values
much lower than $\beta_{cu}$  there is no clear signal for the $\hat{\sigma}_u$
correlation  function.
 The  results shown in tables~\ref{tab_pi1} and~\ref{tab_pi2} were obtained
without including the pseudoscalar contribution in the fit.
Apparently, the pseudoscalar state does not give an important contribution to
correlation functions of type 1, its amplitude is about an order of
magnitude smaller than the one of the scalar state
and $\chi^2$ per degree of freedom
is the same  whether the pseudoscalar state is included in the
fit or not. The
difference in the results from both fits is of the same order as the error
due to the choice of the fit interval.
 Fit intervals were $t_{min}/t_{max}$ = $2/10$.
The $\hat{\sigma}_u$ energies lie for $\beta$ values between
0.15 and 0.17 slightly
below $2m_{uR}$ so that they might be bound in this region.  For larger
$\beta$ they get larger than twice the renormalized fermion mass.
In the parameter region studied, all $\hat{\sigma}_d$ masses are larger than
$2m_{dR}$.  From the simulation results at $\beta \geq 0.075$, one observes
that far in the symmetric phase $\pi_d$ and $\hat{\sigma}_d$
states are degenerate, as one would
expect from restoration of chiral symmetry.

Correlation functions for charged composite states ($\bar{u}d$) have been
calculated for $\beta$ values  around $\beta_{cu}$, $\beta_{cd}$
and for $\beta = 0.12$. For small $\beta$ the correlation functions suffer
from bad noise problems, and the signal appears to fall off with an energy
larger than the inverse
lattice spacing. For large $\beta$ there is a good signal, but
the correlation function falls off with an energy which is
$O(1)$ in lattice units and thus
much  larger than the sum
of the renormalized fermion masses. At $\beta = 0.12$  there is a signal in
the channel considered and
the energy is of the order of $m_{uR}$. However, the renormalized $u$ mass
is large ($O(1)$) at those values of the coupling and
cannot be determined reliably because correlators are very noisy.
Thus no evidence for
 charged bound  states, with small energies  in the limit of large
cut-off, has been found.
\subsection{Renormalization group flows}
In section~\ref{sec:charge} it was discussed that in the \twm\ the cut-off
cannot be removed if the renormalized coupling is kept finite. In such cases
there is always the question if  there are parameter regions
 where physics can be kept fairly constant while the cut-off is
changed. In QED with one set of charges, in general  flow lines of constant
mass  ratios involving composite states cross flow lines
of constant renormalized charge even before the maximal cut-off resulting from
triviality is reached. Only in the sector with very small charges and masses
the flow lines are nearly parallel and physics can  be kept nearly constant.
There are indications that the situation can be improved if four-fermion
interactions are included~\cite{rakow91}. In
the \twm\ the two species of fermions interact with each other only weakly
and one expects the behaviour of flows belonging to the $d$ to be completely
different from the behaviour of flow lines belonging to the $u$.

\begin{figure}[t]
\leavevmode
\vspace{-5cm}
\centerline{
\epsfysize=14cm
\epsfbox{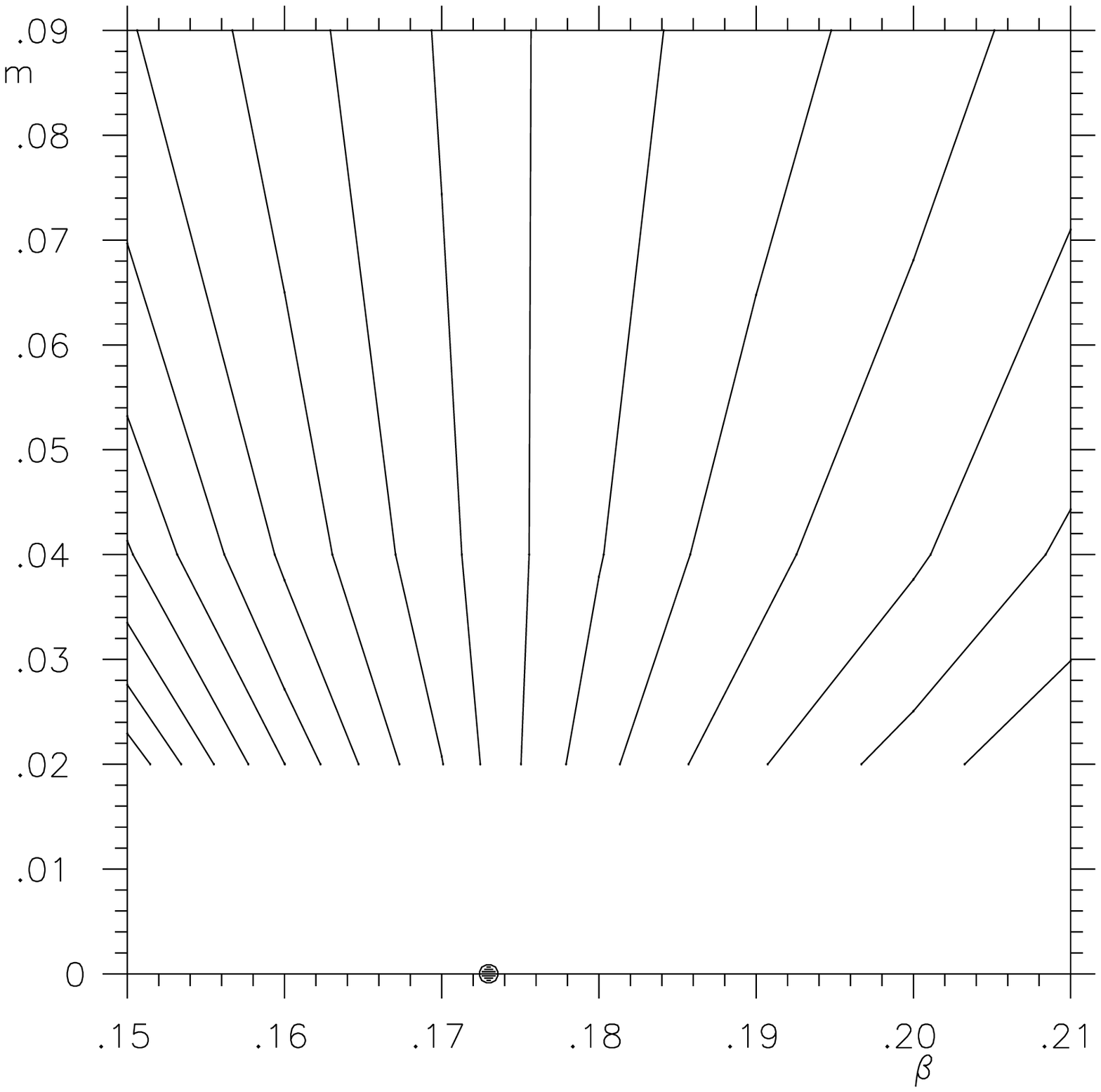}
}
\caption{Lines with constant $m_{uR}/m_{\pi_{u}}$ in the
  $m_u = m_d$ plane. The black dot denotes $\beta_{cu}$.}
\label{fig:mu-pu}
\leavevmode
\vspace{0.5cm}
\centerline{
\epsfysize=8cm
\epsfbox{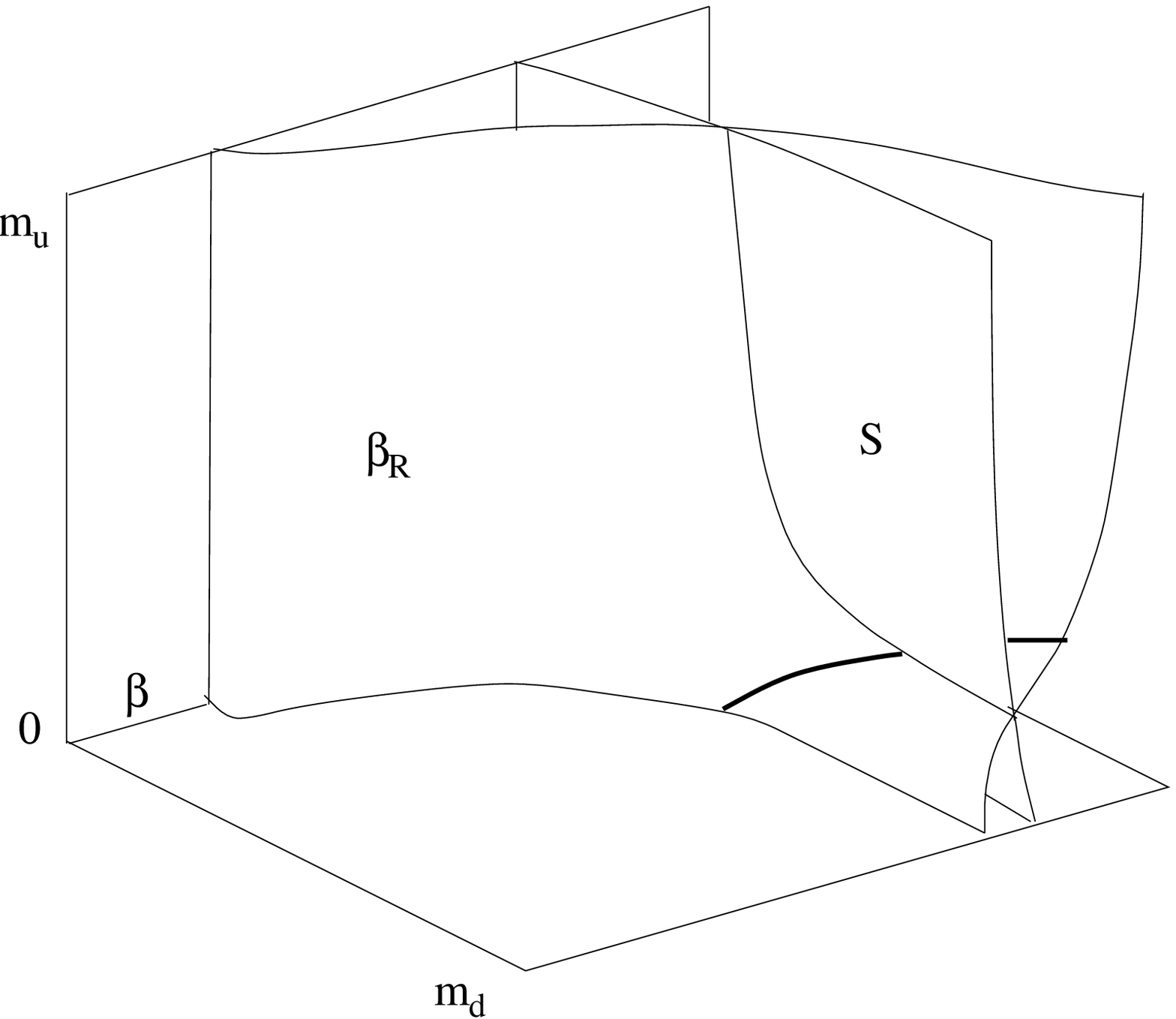}
}
\caption{Sketch of a surface with constant $\beta_R$  and a surface
 with constant $m_{uR}/m_{\pi_{u}}$. The thick black line denotes a constant
fermion mass ratio on the surface with constant $\beta_R$.}
\label{fig:flowsketch}
 \end{figure}

The physically most interesting region in the phase diagram is around $\beta
\simeq \beta_{cu}$ and at small bare masses. Looking at
table~\ref{tab_pi2},
it seems reasonable to assume that energies of composite states of one fermion
only have a very small dependence on the bare mass of the other fermion and
that thus
 the flow lines in the plane $m\equiv m_u = m_d$  can easily
be generalized to the
two dimensional flow surface. To obtain flow lines in the $m_u = m_d$ plane,
an interpolation between the grid of actual simulation results in this plane
is performed. For this, the following dependence of mass ratios $S$ on the
simulation parameters is assumed:
\begin{eqnarray}
 \ln S &=& a + b \beta, \; m \; \mbox{fixed} \label{eq:fl1} \\
\ln S  &=& c + d\ln m, \; \beta \; \mbox{fixed}. \label{eq:fl2}
\end{eqnarray}
The Ansatz in eq. (\ref{eq:fl1}) is motivated by the logarithmic
relation eq. (\ref{eq:mlim1}) between renormalized masses and
the  charge, whereas eq.~(\ref{eq:fl2})  is motivated by
the scaling behaviour expected from the \eqss.
Figure~\ref{fig:mu-pu} shows lines with a constant ratio
$S =  m_{uR}/m_{\pi_{u}}$ in the $m_u = m_d$ plane. $S$ is varied in steps of
0.1 from 0.4 to 2.1. The picture suggests that the lines flow into
$\beta_{cu}$.
Generalizing this to different values of $m_u$ and $m_d$,
one obtains a picture about the flows as shown in
figure~\ref{fig:flowsketch}. Surfaces $S = const$ are expected to end on the
$m_u = 0$ plane at the phase boundary of the $u$. As the thick black line in
figure~\ref{fig:flowsketch} indicates, one generally
cannot keep the ratio of fermion and pion masses constant on a line with
constant fermion mass ratio and renormalized charge, except probably in the
perturbative region.

\begin{figure}[b]
\leavevmode
\vspace{-5cm}
\centerline{
\epsfysize=14cm
\epsfbox{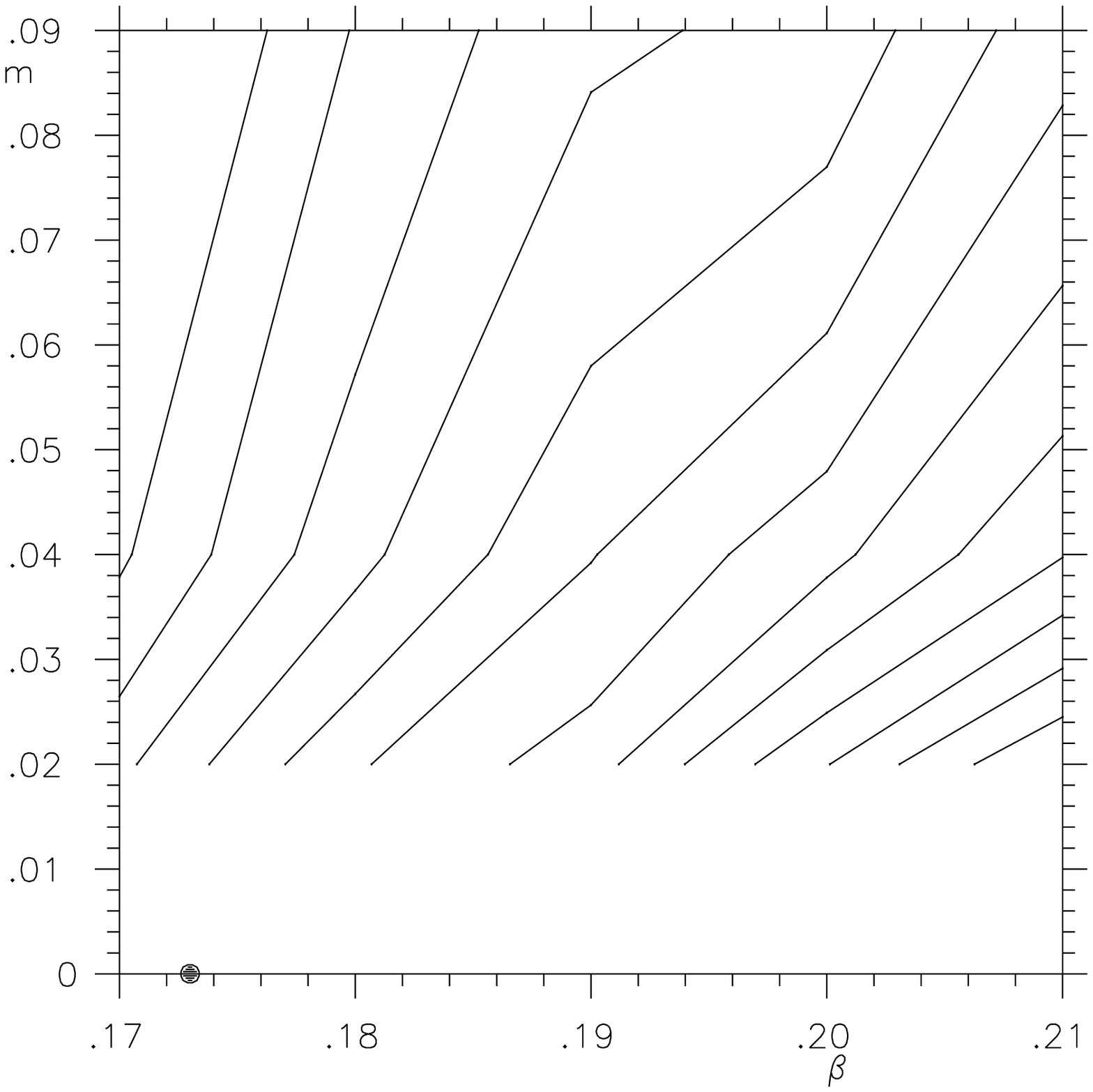}
}
\caption{Lines with constant $m_{uR}/m_{\sigma_{u}}$ in the
  $m_u = m_d$ plane. The black dot denotes $\beta_{cu}$.}
\label{fig:mu-su}
%
\leavevmode
\vspace{-5cm}
\centerline{
\epsfysize=14cm
\epsfbox{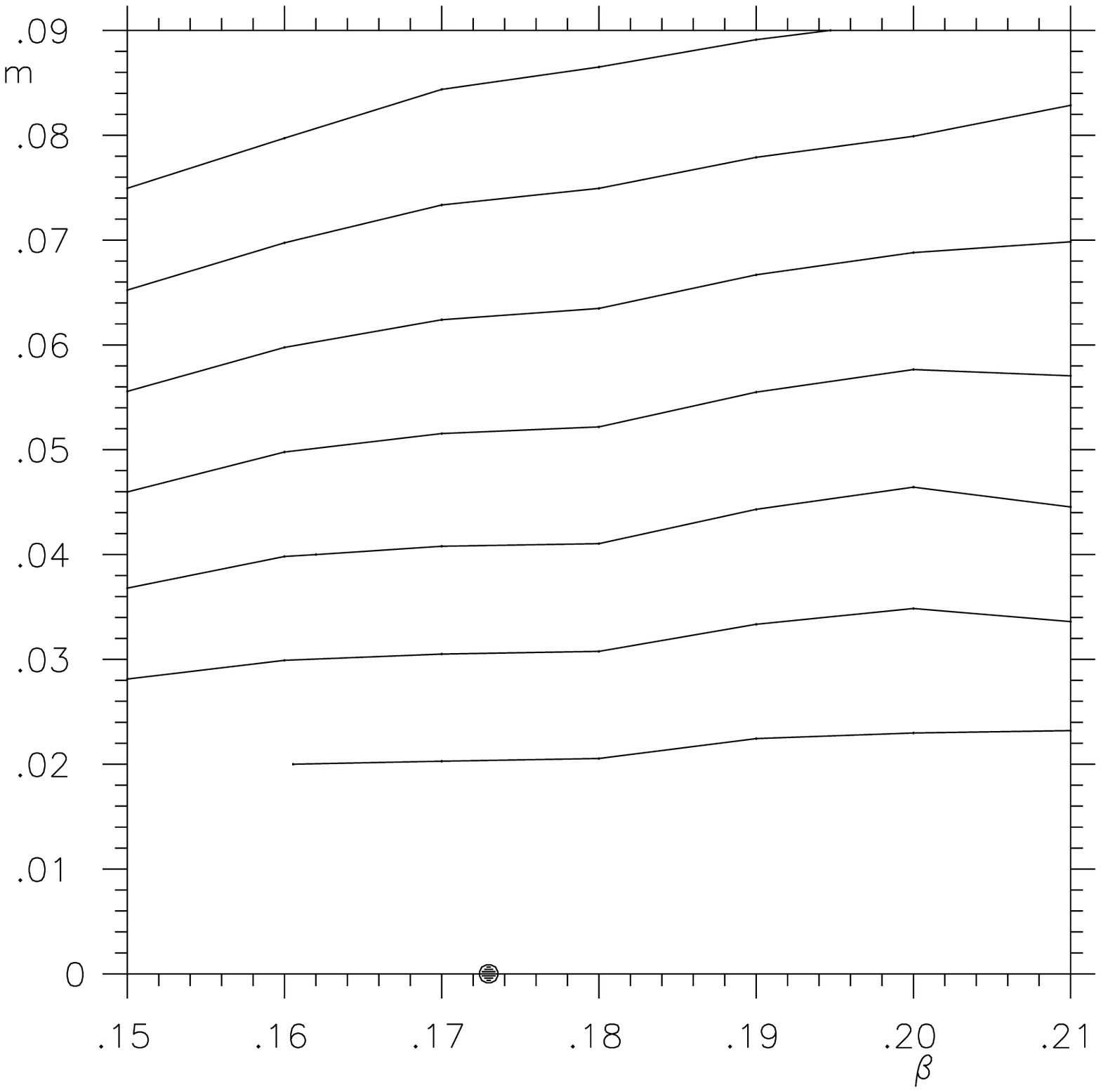}
}
\caption{Lines with constant $m_{dR}/m_{\pi_{d}}$ in the
  $m_u = m_d$ plane. The black dot denotes $\beta_{cu}$.}
\label{fig:md-pd}
 \end{figure}

Figure~\ref{fig:mu-su} shows lines with a constant ratio $S =
m_{uR}/m_{\hat{\sigma}_u}$.  $S$ is varied in steps of 0.025 from 0.3 to 0.6.
The
lines do not flow into $\beta_{cu}$.  One expects that lines of constant
$\beta_R$ and fermion mass ratio will also in general not lie on the surface
one obtains from generalizing $S$ to the three-dimensional parameter space.
However, for very small couplings and masses it seems that flow lines with
constant coupling and fermion mass ratio follow surfaces with constant mass
ratios closely and renormalizability  is essentially restored.
Lines with constant $S$ = $m_{dR}$/\mpd\ are shown in figure~\ref{fig:md-pd}.
$S$ is varied in steps of 0.0125 from 0.025 to 0.1. The lines show that this
mass ratio is for large couplings fairly independent of $m_d$.

\clearpage
\section{Conclusions}
In this paper a lattice study of non-compact QED with two sets of
staggered fermions with charges 1 ($u$)  and $-1/2$ ($d$)
(`two-charge model'), is presented. The phase diagram is obtained from
the chiral condensates.  They can be
described by a fit with equations of state of an $O(2)$ symmetric linear
sigma model with logarithmic corrections to the mean-field equations.
Chiral symmetry breaking occurs at different values of the bare coupling for
both fermions, for the $u$ fermion at $\beta_{cu}=0.173(1)$ and for the $d$
fermion  at $\beta_{cd}=0.047(1)$.  The most interesting
 candidate for a continuum limit of the model is at $\beta_{cu}$, with
$m_u$ = $m_d$ = 0. This is the end point of the line on the $\beta$ axis where
  renormalized  masses of both fermions are zero in units of the cut-off.
There are indications that for smaller $\beta$ the renormalized $d$ mass
can go to zero, while the renormalized $u$ mass  is finite. If $\beta$ is
lowered past $\beta_{cd}$, both renormalizd masses are always finite.

The renormalized coupling has been determined and  found to  be compatible
with perturbation theory. Other effects to the charge renormalization like
possible charged bound states seem not to give a noticeable contribution.
The agreement with perturbation theory indicates that the
renormalized charge of all fermions vanishes even if only one becomes massless.
An estimate for the validity bound of the \twm\ was obtained and
generalized to all charged fermions in
the Standard Model. Including all known charged fermions one
gets an upper bound of
 $\alpha_R \lesssim 1/80$ if one assumes  QED to be valid up
to the Planck scale.

Study of composite states ($\bar{u}u$ and $\bar{d}d$) has shown that in the
neighbourhood of the physically interesting point $\beta_{cu}$ only the
$\bar{u}u$ states are bound. Masses of $\bar{d}d$ states are of $O(1)$ in
lattice units in this region of the phase diagram.  It appears that due to
the shape of renormalization group flows of mass ratios one cannot keep
physics constant even approximately in the investigated parameter region.
The theory seems to
 become inconsistent already at scales which are lower than the
cut-off due to triviality. However there are indications that
renormalizability is approximately restored
in the perturbative region. The situation that the theory is in general
 not renormalizable
 may be improved by including other  operators into the action.
\section*{Acknowledgements}
I am grateful to my PhD supervisor G. Schierholz for many inspiring
suggestions  and would like to thank
 R. Horsley, M. G\"ockeler, P. Rakow and H. St\"uben for valuable
discussions. The numerical calculations were performed on the Cray Y-MP at
HLRZ J\"ulich and the Alliant FX2816 at GMD, St. Augustin, and I would like
to thank both computer centers for their support. This work was partly
supported by SHEFC and the EU under contract EC CHRX-CT92-0051.

\clearpage
\section*{Tables}
\vspace{3cm}
\begin{table}[h]
    \begin{center}
        \begin{tabular}{||c|c|c|l|l||}
           \hline
           \multicolumn{1}{||c|}{$\beta$} &
           \multicolumn{1}{c|}{$m_{u}$} &
           \multicolumn{1}{c|}{$m_{d}$} &
           \multicolumn{1}{|c|}{$<\bar{\chi}_{u}\chi_{u}>$} &
           \multicolumn{1}{|c||}{$<\bar{\chi}_{d}\chi_{d}>$} \\
           \hline
           \hline
 $0.00$ &    $0.09$ & $0.09$ & $0.6312(11)   $ & $0.6323(11) $  \\
    \hline
 $0.04$ & $0.04$ & $0.04$ & $0.6346(17)   $ & $0.4036(13)     $ \\
    \hline
 $0.05$ & $0.02$ & $0.02$ & $0.6302(23)   $ & $0.1881(15)    $ \\
 $0.05$ &$0.04 $ &$0.04$ & $0.6295(17) $ & $  0.2515(10)      $ \\
 $0.05$ &$0.09 $ &$0.09$ & $0.6258(11) $ & $  0.3339(8) $        \\
    \hline
 $0.06$ &  $  0.02  $&$   0.04 $ & $   0.6180(24) $ &$   0.1392(6) $ \\
 $0.06$ &  $  0.04  $&$   0.04 $ & $   0.6243(17) $ &$   0.1397(5) $ \\
    \hline
 $0.075$ &  $  0.02  $&$   0.02 $ & $   0.6067(23) $ &$   0.03988(10) $ \\
 $0.075$ &  $  0.04  $&$   0.04 $ & $   0.6073(16) $ &$   0.07898(22) $ \\
 $0.075$ &  $  0.09  $&$   0.09 $ & $   0.6061(11) $ &$   0.1662(3) $ \\
    \hline
 $0.10$ &  $  0.02  $&$   0.02 $ & $   0.5569(22) $ &$   0.02602(4)  $ \\
 $0.10$ &  $  0.04  $&$   0.04 $ & $   0.5640(16) $ &$   0.05380(9)  $ \\
 $0.10$ &  $  0.09  $&$   0.09 $ & $   0.5664(10) $ &$   0.1148(1)  $ \\
   \hline
 $0.12$ &  $  0.02  $&$   0.04 $ & $   0.4970(19) $ &$   0.04357(5)  $ \\
   \hline
     \end{tabular}
 \end{center}
 \caption[xxx]{Chiral condensates on the $
   8^{3}\times 12$ lattice, region $\beta\sim\beta_{cd}$.}
 \label{table1}

\end{table}
 \begin{table}[p]
    \begin{center}
        \begin{tabular}{||c|c|c|l|l||}
           \hline
           \multicolumn{1}{||c|}{$\beta$} &
           \multicolumn{1}{c|}{$m_{u}$} &
           \multicolumn{1}{c|}{$m_{d}$} &
           \multicolumn{1}{|c|}{$<\bar{\chi}_{u}\chi_{u}>$} &
           \multicolumn{1}{|c||}{$<\bar{\chi}_{d}\chi_{d}>$} \\
           \hline
    \hline
 $0.14$ &  $  0.02  $&$   0.02 $ & $   0.4222(19) $ &$   0.01951(3)  $ \\
    \hline
 $0.15$ &  $  0.02  $&$   0.02 $ & $   0.3741(18) $ &$   0.01880(2)  $ \\
 $0.15$ &  $  0.04  $&$   0.04 $ & $   0.4031(12) $ &$   0.03810(5)  $ \\
 $0.15$ &  $  0.09  $&$   0.09 $ & $   0.4469(09) $ &$   0.08451(9)  $ \\
    \hline
 $0.16$ &  $  0.02  $&$   0.02 $ & $   0.3236(17) $ &$   0.01813(2)  $ \\
 $0.16$ &  $  0.02  $&$   0.04 $ & $   0.3316(16) $ &$   0.03595(3)  $ \\
 $0.16$ &  $  0.04  $&$   0.02 $ & $   0.3656(12) $ &$   0.01815(2)  $ \\
 $0.16$ &  $  0.04  $&$   0.04 $ & $   0.3629(12) $ &$   0.03611(3)  $ \\
 $0.16$ &  $  0.09  $&$   0.09 $ & $   0.4168(8)  $ &$   0.08146(8)  $ \\
    \hline
 $0.17$ &  $  0.02  $&$   0.02 $ & $   0.2742(15) $ &$   0.01772(2)  $ \\
 $0.17$ &  $   0.04 $&$    0.04$ & $    0.3238(12)$ &$    0.03539(4) $ \\
 $0.17$ &  $   0.09 $&$    0.09$ & $    0.3908(8) $ &$    0.07871(7) $ \\
    \hline
 $0.18$ &  $   0.02 $&$    0.02$ & $    0.2167(14)$ &$    0.01741(2) $ \\
 $0.18$ &  $   0.02 $&$    0.04$ & $    0.2227(15)$ &$    0.03435(3) $ \\
 $0.18$ &  $   0.02 $&$    0.09$ & $    0.2255(14)$ &$    0.07602(7) $ \\
 $0.18$ &  $  0.04  $&$   0.02 $ & $   0.2850(11) $ &$   0.01725(2)  $ \\
 $0.18$ &  $  0.04  $&$   0.04 $ & $   0.2854(11) $ &$   0.03524(5)  $ \\
 $0.18$ &  $  0.04  $&$   0.09 $ & $   0.2844(12) $ &$   0.07702(8)  $ \\
 $0.18$ &  $  0.09  $&$   0.02 $ & $   0.3611(8)  $ &$   0.01737(2)  $ \\
 $0.18$ &  $  0.09  $&$   0.04 $ & $   0.3618(8)  $ &$   0.03473(3)  $ \\
 $0.18$ &  $  0.09  $&$   0.09 $ & $   0.3621(8)  $ &$   0.07716(7)  $ \\
           \hline
 $0.19$ &  $  0.02  $&$   0.02 $ & $   0.1870(15) $ &$   0.01669(1)  $ \\
 $0.19$ &  $  0.04  $&$   0.04 $ & $   0.2444(10) $ &$   0.03370(4)  $ \\
 $0.19$ &  $  0.09  $&$   0.09 $ & $   0.3372(7)  $ &$   0.07502(6)  $ \\
           \hline
 $0.20$ &  $  0.02  $&$   0.02 $ & $   0.1416(10) $ &$   0.01649(1)  $ \\
 $0.20$ &  $  0.02  $&$   0.16 $ & $   0.1462(15) $ &$   0.1263(1) $ \\
 $0.20$ &  $  0.04  $&$   0.04 $ & $   0.2167(9)  $ &$   0.03283(3)  $ \\
 $0.20$ &  $  0.09  $&$   0.09 $ & $   0.3108(8)  $ &$   0.07420(6)  $ \\
 $0.20$ &  $  0.16  $&$   0.02 $ & $   0.3785(6)  $ &$   0.01677(1)  $ \\
           \hline
 $0.21$ &  $  0.02  $&$   0.02 $ & $   0.1143(7)  $ &$   0.01624(1)  $ \\
 $0.21$ &  $  0.02  $&$   0.16 $ & $   0.1072(7)  $ &$   0.1254(1) $ \\
 $0.21$ &  $  0.04  $&$   0.04 $ & $   0.1838(8)  $ &$   0.03331(6)  $ \\
 $0.21$ &  $  0.09  $&$   0.09 $ & $   0.2882(7)  $ &$   0.07280(6)  $ \\
 $0.21$ &  $  0.16  $&$   0.02 $ & $   0.3605(5)  $ &$   0.01660(1)  $ \\
           \hline
 $0.22$ &  $  0.04  $&$   0.04 $ & $   0.1601(7)  $ &$   0.03222(3)  $ \\
 $0.22$ &  $  0.09  $&$    0.09$ & $    0.2693(6) $ &$    0.07162(6) $ \\
           \hline
      \end{tabular}
 \end{center}
 \caption[xxx]{Chiral condensates on the
   $8^{3}\times 12$ lattice, region $\beta\sim\beta_{cu}$.}
 \label{table2}

\end{table}

\begin{table}[tb]
    \begin{center}
        \begin{tabular}{||l|l|l|l||}
           \hline
           \hline
$\beta_{cu}    $ & $0.1729(2)   $ & $\beta_{cd}     $ & $0.04675(7) $ \\
$\tau_u^{(1)}  $ & $-1.78(1)    $ & $\tau_d^{(1)}   $ & $-1.469(9) $ \\
$\tau_u^{(3)}  $ & $0.48(3)     $ & $               $ & $ $ \\
$\theta_u^{(0)}$ & $2.61(2)   $ &   $\theta_d^{(0)} $ & $2.41(2) $ \\
$\theta_u^{(1)}$ & $-1.3(1)  $ &   $\theta_d^{(1)} $ & $-1.42(2) $ \\
$\theta_u^{(3)}$ & $-5.7(6)   $ &   $               $ & $ $ \\
$p_u           $ & $0.588(6)  $ &   $p_d            $ & $0.074(3)  $ \\
\hline
    \end{tabular}
 \end{center}
 \caption[xxx]{Fit parameters for the equation of state with logarithmic
corrections.}
 \label{tab:mf-logs}

\end{table}
\begin{table}
    \begin{center}
        \begin{tabular}{||c|c|c|l|l|l||}
           \hline
           \multicolumn{1}{||c|}{$\beta$} &
           \multicolumn{1}{c|}{$m_{u}$} &
           \multicolumn{1}{c|}{$m_{d}$} &
           \multicolumn{1}{|c|}{$m_{uR}$} &
           \multicolumn{1}{c|}{$m_{dR}$} &
           \multicolumn{1}{|c||}{$\beta_R$} \\
          \hline
   $0.05$ & $0.02$ & $0.02$ & $5.000(675)$ & $0.326(7)$ & $ 0.073(1) $  \\
   $0.05$ & $0.04$ & $0.04$ & $4.935(143)$ & $0.446(4)$ & $ 0.069(1) $  \\
   $0.05$ & $0.09$ & $0.09$ & $5.646(604)$ & $0.644(7)$ & $ 0.063(1) $  \\
\hline
   $0.06$ & $0.02$ & $0.04$ & $2.872(393)$ & $0.258(4)$ & $ 0.087(1) $  \\
   $0.06$ & $0.04$ & $0.04$ & $2.573(382)$ & $0.252(3)$ & $ 0.088(1) $  \\
\hline
   $0.075$ & $0.02$ & $0.02$ & $2.018(143)$ & $0.085(1)$ & $ 0.123(1) $  \\
   $0.075$ & $0.04$ & $0.04$ & $1.993(210)$ & $0.149(2)$ & $ 0.116(1) $  \\
   $0.075$ & $0.09$ & $0.09$ & $2.912(126)$ & $0.295(2)$ & $ 0.104(1) $  \\
\hline
   $0.10$ & $0.02$ & $0.02$ & $1.518(074)$ & $0.050(1)$ & $ 0.163(2) $  \\
   $0.10$ & $0.04$ & $0.04$ & $1.864(120)$ & $0.090(1)$ & $ 0.154(1) $  \\
   $0.10$ & $0.09$ & $0.09$ & $1.662(92) $ & $0.209(2)$ & $ 0.139(1) $  \\
\hline
   $0.12$ & $0.02$ & $0.04$ & $1.124(37) $ & $0.072(1)$ & \\
\hline
    \end{tabular}
 \end{center}
 \caption[xxx]{Renormalized fermion masses and coupling,
 region $\beta\sim\beta_{cd}$.}
 \label{tab_mr1}
\end{table}
\clearpage
\begin{table}
    \begin{center}
        \begin{tabular}{||c|c|c|l|l|l||}
           \hline
           \multicolumn{1}{||c|}{$\beta$} &
           \multicolumn{1}{c|}{$m_{u}$} &
           \multicolumn{1}{c|}{$m_{d}$} &
           \multicolumn{1}{|c|}{$m_{uR}$} &
           \multicolumn{1}{c|}{$m_{dR}$} &
          \multicolumn{1}{|c||}{$\beta_R$} \\
           \hline
   $0.14$ & $0.02$ & $0.02$ & $0.917(28) $ & $0.034(1)$ & $ 0.238(3) $  \\
\hline
   $0.15$ & $0.02$ & $0.02$ & $0.753(13) $ & $0.032(1)$ & $ 0.257(2) $  \\
   $0.15$ & $0.04$ & $0.04$ & $0.858(9)  $ & $0.065(1)$ & $ 0.239(2) $  \\
   $0.15$ & $0.09$ & $0.09$ & $1.023(13) $ & $0.148(1)$ & $ 0.215(2) $  \\
\hline
   $0.16$ & $0.02$ & $0.02$ & $0.609(7)  $ & $0.031(1)$ & $ 0.276(3) $  \\
   $0.16$ & $0.02$ & $0.04$ & $0.617(8)  $ & $0.063(1)$ & $ 0.264(3) $  \\
   $0.16$ & $0.04$ & $0.02$ & $0.727(9)  $ & $0.033(1)$ & $ 0.272(2) $  \\
   $0.16$ & $0.04$ & $0.04$ & $0.714(10) $ & $0.064(1)$ & $ 0.258(2) $  \\
   $0.16$ & $0.09$ & $0.09$ & $0.918(9)  $ & $0.143(1)$ & $ 0.232(2) $  \\
\hline
   $0.17$ & $0.02$ & $0.02$ & $0.485(7)  $ & $0.030(1)$ & $ 0.304(3) $  \\
   $0.17$ & $0.04$ & $0.04$ & $0.607(8)  $ & $0.061(1)$ & $ 0.274(3) $  \\
   $0.17$ & $0.09$ & $0.09$ & $0.815(5)  $ & $0.139(1)$ & $ 0.250(2) $  \\
\hline
   $0.18$ & $0.02$ & $0.02$ & $0.366(7)  $ & $0.029(1)$ & $ 0.329(3) $  \\
   $0.18$ & $0.02$ & $0.04$ & $0.384(5)  $ & $0.058(1)$ & $ 0.321(3) $  \\
   $0.18$ & $0.02$ & $0.09$ & $0.379(4)  $ & $0.130(1)$ & $ 0.305(3) $  \\
   $0.18$ & $0.04$ & $0.02$ & $0.506(5)  $ & $0.029(1)$ & $ 0.315(3) $  \\
   $0.18$ & $0.04$ & $0.04$ & $0.512(6)  $ & $0.058(1)$ & $ 0.303(3) $  \\
   $0.18$ & $0.04$ & $0.09$ & $0.512(5)  $ & $0.130(1)$ & $ 0.288(3) $  \\
   $0.18$ & $0.09$ & $0.02$ & $0.720(4)  $ & $0.030(1)$ & $ 0.291(2) $  \\
   $0.18$ & $0.09$ & $0.04$ & $0.715(4)  $ & $0.060(1)$ & $ 0.278(2) $  \\
   $0.18$ & $0.09$ & $0.09$ & $0.720(6)  $ & $0.133(1)$ & $ 0.268(2) $  \\
\hline
   $0.19$ & $0.02$ & $0.02$ & $0.319(6)  $ & $0.029(1)$ & $ 0.357(3) $  \\
   $0.19$ & $0.04$ & $0.04$ & $0.425(4)  $ & $0.057(1)$ & $ 0.321(3) $  \\
   $0.19$ & $0.09$ & $0.09$ & $0.658(4)  $ & $0.134(1)$ & $ 0.279(2) $  \\
\hline
   $0.20$ & $0.02$ & $0.02$ & $0.229(7)  $ & $0.028(1)$ & $ 0.388(3) $  \\
   $0.20$ & $0.02$ & $0.16$ & $0.245(4)  $ & $0.219(1)$ & $ 0.351(3) $  \\
   $0.20$ & $0.04$ & $0.04$ & $0.372(5)  $ & $0.056(1)$ & $ 0.342(3) $  \\
   $0.20$ & $0.09$ & $0.09$ & $0.596(6)  $ & $0.128(1)$ & $ 0.299(2) $  \\
   $0.20$ & $0.16$ & $0.02$ & $0.807(6)  $ & $0.030(1)$ & $ 0.303(2) $  \\
\hline
   $0.21$ & $0.02$ & $0.02$ & $0.182(4)  $ & $0.027(1)$ & $ 0.415(2) $  \\
   $0.21$ & $0.02$ & $0.16$ & $0.171(2)  $ & $0.213(1)$ & $ 0.351(3) $  \\
   $0.21$ & $0.04$ & $0.04$ & $0.311(3)  $ & $0.055(1)$ & $ 0.367(3) $  \\
   $0.21$ & $0.09$ & $0.09$ & $0.528(5)  $ & $0.125(1)$ & $ 0.315(2) $  \\
   $0.21$ & $0.16$ & $0.02$ & $0.735(7)  $ & $0.029(1)$  & $ 0.318(2) $ \\
\hline
   $0.22$ & $0.04$ & $0.04$ & $0.268(4)  $ & $0.054(1)$ & $ 0.386(3) $  \\
   $0.22$ & $0.09$ & $0.09$ & $0.491(3)  $ & $0.124(1)$ & $ 0.328(2) $  \\
\hline
    \end{tabular}
 \end{center}
 \caption[xxx]{Renormalized fermion masses and coupling,
 region $\beta\sim\beta_{cu}$.}
 \label{tab_mr2}

\end{table}
\begin{table}
    \begin{center}
        \begin{tabular}{||l|l|l|l||}
           \hline
$P_1$ & $ 1.54(1) $ & $ Q_1 $ & $ 1.726(3) $ \\
$P_2$ & $ 3.5(1) $ & $ Q_2 $ &  \\
\hline
    \end{tabular}
 \end{center}
 \caption[xxx]{Fit parameters for the renormalized fermion masses.}
 \label{tab:mrfit}
\end{table}
\begin{table}
\begin{center}
\begin{tabular}{|c|l|c|c|}
\hline
$i$ & $s_{i}(\vec{x},t)$     & quantum numbers    & continuum states \\
\hline
\hline
1   & $(-1)^t $                       & $0_s^{++}$ & $\hat{\sigma}$ \\
    &                          & $0_a^{-+}$ & $\pi'$ \\
\hline
2   & $(-1)^{x_1+x_2+x_3+t}$     & $0_a^{+-}$ & $-$ \\
    &                          & $0_a^{-+}$ & $\pi$ \\
\hline
\end{tabular}
\end{center}
\caption{Sign factors for the meson operators. The $\sigma$ particle is
provided with a hat to avoid confusion with the chiral condensates
introduced earlier.}
\label{tab:localmes}
\end{table}

\begin{table}
    \begin{center}
        \begin{tabular}{||c|c|c|l|l|l||}
           \hline
           \multicolumn{1}{||c|}{$\beta$} &
           \multicolumn{1}{c|}{$m_{u}$} &
           \multicolumn{1}{c|}{$m_{d}$} &
           \multicolumn{1}{|c|}{$m_{\pi_u}$} &
           \multicolumn{1}{|c|}{$m_{\pi_d}$} &
           \multicolumn{1}{c||}{$m_{\sigma_d}$} \\
           \hline
           \hline
$0.05$& $ 0.02$ & $0.02$ & $ 0.305(1)$  &  $0.431(1)$ & $ 0.714(12)$\\
$0.05$& $ 0.04$ & $0.04$ & $ 0.431(0)$  &  $0.556(1)$ & $ 0.950(21)$\\
$0.05$& $ 0.09$ & $0.09$ & $ 0.648(0)$  &  $0.783(1)$ & $ 1.219(16)$\\
           \hline
$0.06$ &$ 0.02$&$  0.04$ &  $0.307(1)$  &  $0.694(2)$ & $0.798(3)$  \\
$0.06$ &$ 0.04$&$  0.04$ &  $0.433(1)$  &  $0.695(2)$ & $0.794(3)$  \\
           \hline
$0.075$ &$ 0.02$&$  0.02$ &  $0.311(1)$  &  $0.937(2)$& $ 0.934(2) $ \\
$0.075$ &$ 0.04$&$  0.04$ &  $0.438(1)$  &  $0.933(2)$& $ 0.941(2) $ \\
$0.075$ &$ 0.09$&$  0.09$ &  $0.657(0)$  &  $0.965(1)$& $ 1.045(2) $ \\
           \hline
$0.10$ & $0.02$ & $0.02$  & $0.319(1)$  &  $1.124(3)$ & $ 1.126(3) $\\
$0.10$ & $0.04$ & $0.04$  & $0.450(1)$  &  $1.042(3)$ & $ 1.052(3) $\\
$0.10$ & $0.09$ & $0.09$  & $0.671(1)$  &  $1.142(3)$ & $ 1.173(3) $\\
\hline
\hline
    \end{tabular}
 \end{center}
 \caption[xxx]{Energies of neutral  composite states,
 region $\beta\sim\beta_{cd}$.}
 \label{tab_pi1}
\end{table}
\begin{table}
    \begin{center}
        \begin{tabular}{||c|c|c|l|l|l|l||}
           \hline
           \multicolumn{1}{||c|}{$\beta$} &
           \multicolumn{1}{c|}{$m_{u}$} &
           \multicolumn{1}{c|}{$m_{d}$} &
           \multicolumn{1}{|c|}{$m_{\pi_u}$} &
           \multicolumn{1}{c||}{$m_{\pi_d}$} &
           \multicolumn{1}{|c|}{$m_{\sigma_u}$} &
           \multicolumn{1}{c||}{$m_{\sigma_d}$} \\
           \hline
           \hline
$0.15$ & $0.02$ & $0.02$  & $0.361(1)$  &  $1.229(2)$ & $ 0.822(130)$& $
1.229(2) $ \\
$0.15$ & $0.04$ & $0.04$  & $0.502(1)$  &  $1.189(2)$ & $ 1.216(99) $& $
1.191(2) $ \\
$0.15$ & $0.09$ & $0.09$  & $0.723(1)$  &  $1.241(2)$ & $ 1.332(94) $& $
1.255(2) $ \\
           \hline
$0.16$& $ 0.02$& $ 0.02$ &  $0.381(1)$ &   $1.239(2)$ & $ 0.847(53) $& $
1.240(2) $ \\
$0.16$& $ 0.02$& $ 0.04$ &  $0.379(1)$ &   $1.261(2)$ & $ 0.769(52) $& $
1.262(2) $ \\
$0.16$& $ 0.04$& $ 0.02$ &  $0.515(1)$ &   $1.264(2)$ & $ 1.161(68) $& $
1.264(2) $ \\
$0.16$& $ 0.04$&  $0.04$ &  $0.514(1)$ &   $1.274(2)$ & $ 1.083(55) $& $
1.275(2) $ \\
$0.16$& $ 0.09$&  $0.09$ &  $0.737(1)$ &   $1.267(2)$ & $ 1.320(55) $& $
1.279(2) $ \\
           \hline
$0.17$&  $0.02$ & $0.02$  & $0.407(1)$  &  $1.217(2)$ & $ 0.881(33) $& $
1.217(2) $ \\
$0.17$&  $0.04$ & $0.04$  & $0.539(1)$  &  $1.243(2)$ & $ 1.010(33) $& $
1.244(2) $ \\
$0.17$&  $0.09$ & $0.09$  & $0.752(1)$  &  $1.307(2)$ & $ 1.266(36) $& $
1.318(2) $ \\
           \hline
$0.18$ & $0.02$ & $0.02$ &  $0.444(1)$  &  $1.192(2)$ & $ 0.774(15) $& $
1.193(2) $ \\
$0.18$ & $0.02$ & $0.04$ &  $0.431(1)$  &  $1.242(2)$ & $ 0.762(15) $& $
1.243(2) $ \\
$0.18$ & $0.02$ & $0.09$ &  $0.436(1)$  &  $1.280(2)$ & $ 0.755(14) $& $
1.287(2) $ \\
$0.18$&  $0.04$ & $0.02$ &  $0.558(1)$  &  $1.274(2)$ & $ 1.003(19) $& $
1.274(2) $  \\
$0.18$&  $0.04$ & $0.04$ &  $0.563(1)$  &  $1.188(3)$ & $ 0.958(17) $& $
1.184(3) $  \\
$0.18$&  $0.04$ & $0.09$ &  $0.563(1)$  &  $1.241(2)$ & $ 0.946(16) $& $
1.254(2) $  \\
$0.18$ & $0.09$ & $0.02$ &  $0.767(1)$  &  $1.283(2)$ & $ 1.266(24) $& $
1.277(2) $ \\
$0.18$ & $0.09$ & $0.04$ &  $0.768(1)$  &  $1.276(2)$ & $ 1.252(23) $& $
1.284(2) $ \\
$0.18$ & $0.09$ & $0.09$ &  $0.769(1)$  &  $1.282(2)$ & $ 1.256(23) $& $
1.293(2) $ \\
           \hline
$0.19$&  $0.02$ & $0.02$ &  $0.444(1)$  &  $1.305(2)$ & $ 0.734(11) $& $
1.305(2) $ \\
$0.19$&  $0.04$ & $0.04$ &  $0.588(1)$  &  $1.262(2)$ & $ 0.903(9)  $& $
1.260(2) $ \\
$0.19$&  $0.09$ & $0.09$ &  $0.776(1)$  &  $1.327(2)$ & $ 1.246(18) $& $
1.334(2) $ \\
           \hline
$0.20$&  $0.02$ & $0.02$ &  $0.509(2)$  &  $1.282(2)$ & $ 0.655(4)  $& $
1.281(2) $ \\
$0.20$&  $0.02$ & $0.16$ &  $0.496(2)$  &  $1.357(2)$ & $ 0.647(5)  $& $
1.390(2) $ \\
$0.20$&  $0.04$ & $0.04$ &  $0.600(1)$  &  $1.306(2)$ & $ 0.856(73) $& $
1.307(2) $ \\
$0.20$&  $0.09$&  $0.09$  & $0.802(1)$  &  $1.294(2)$ & $ 1.159(10) $& $
1.302(2) $ \\
$0.20$&  $0.16$&  $0.02$  & $0.995(1)$  &  $1.314(2)$ & $ 1.432(17) $& $
1.314(2) $ \\
           \hline
$0.21$ & $0.02$&  $0.02$ &  $0.579(2)$  &  $1.271(2)$ & $ 0.659(2)  $& $
1.271(2) $ \\
$0.21$ & $0.02$&  $0.16$ &  $0.592(2)$  &  $1.317(2)$ & $ 0.667(3)  $& $
1.366(2) $ \\
$0.21$ & $0.04$&  $0.04$ &  $0.645(1)$  &  $1.212(3)$ & $ 0.824(4)  $& $
1.205(3) $ \\
$0.21$ & $0.09$&  $0.09$ &  $0.806(1)$  &  $1.326(2)$ & $ 1.154(9)  $& $
1.332(2) $ \\
$0.21$ & $0.16$&  $0.02$ &  $1.009(1)$  &  $1.285(2)$ & $ 1.395(13) $& $
1.286(2) $ \\
           \hline
$0.22$ & $0.04$&  $0.04$ &  $0.679(2)$  &  $1.282(2)$ & $ 0.817(4)  $& $
1.281(2) $ \\
$0.22$ & $0.09$&  $0.09$ &  $0.818(1)$  &  $1.336(2)$ & $ 1.134(7)  $& $
1.343(2) $ \\
\hline
    \end{tabular}
 \end{center}
 \caption[xxx]{Energies of neutral composite states,
 region $\beta\sim\beta_{cu}$.}
 \label{tab_pi2}
\end{table}


\begin{thebibliography}{99}
\bibitem{landau} L. D. Landau, A. A. Abrikosov and I. M. Khalatnikov, Dokl.
Akad. Nauk {\bf 95} (1954) 177; \\
L. D. Landau and I. Ya. Pomeranchuk, Dokl. Akad. Nauk. {\bf
  102} (1955) 489; \\
L. D. Landau, in {\em  Niels Bohr and the Development of Physics}, ed. W. Pauli
(Pergamon, London, 1955); \\
L. D. Landau, A. A. Abrikosov and I. M. Khalatnikov, Nuovo Cimento, Supplement
{\bf 3} (1956) 80.
\bibitem{feynman85} R. P. Feynman, in {\em QED, the Strange Theory of Light and
  Matter} (Princeton University Press, Princeton, 1985), chapter 4.
\bibitem{miransky}
V. A. Miransky, Nuovo Cim. {\bf 90A} (1985) 149;
Sov. Phys. JETP {\bf 61} (1985) 905;\\
P. I. Fomin, V. P. Gusynin, V. A. Miransky and Yu. A. Sitenko,
Riv. Nuovo Cim. {\bf 6} (1983) 1.
\bibitem{bartho84} J. Bartholomew, S. H. Shenker, J. Sloan, J. Kogut,
M. Stone, H. W. Wyld, J.~Shigemitsu and D. K. Sinclair, Nucl. Phys.
{\bf B230} [FS10] (1984) 222.
\bibitem{salm90} M. Salmhofer and E. Seiler, Lett. Math. Phys. {\bf 20}
(1990).
\bibitem{kondo91a} K. Kondo and H. Nakatani, Nucl. Phys. {\bf B351} (1991) 236.
\bibitem{anom-dim} T. Appelquist, D. Karabali and L. C. R.
Wijewardhana, Phys. Rev. Lett. {\bf 57} (1986) 957; \\
K. Yamawaki, M. Bando and K. Matumoto, Phys. Rev.
Lett. {\bf 56} (1986) 1335; \\
M. Bando, T. Morozumi, H. So and K. Yamawaki,
Phys. Rev. Lett. {\bf 59} (1987) 389.
\bibitem{tanabashi91} M. Tanabashi, in {\em Proceedings of the 1991 Nagoya
  Spring School on   Dynamical Symmetry Breaking}, Nagoya, Japan, p. 336.
\bibitem{kogu88} J. B. Kogut, E. Dagotto and A. Koci\'{c}, Phys. Rev. Lett.
{\bf 60} (1988) 772.
\bibitem{kogu89} J. B. Kogut, E. Dagotto and A. Koci\'{c}, Nucl. Phys.
 {\bf B317} (1989) 253; {\em ibid.} {\bf B317} (1989) 271.
\bibitem{kogu90} E. Dagotto, A. Koci\'{c} and J. B. Kogut, Nucl. Phys. {\bf
B331}
(1990) 500.
\bibitem{Kocic93} A. Koci\'{c}, J. B. Kogut, M.-P. Lombardo and K. C.~Wang,
Nucl.
Phys. {\bf B397} (1993) 451.
\bibitem{boot89} S. P. Booth, R. D. Kenway, B. J. Pendleton, Phys. Lett.
{\bf 228B} (1989) 115.
\bibitem{schi90} M. G\"ockeler, R. Horsley, E. Laermann, P. Rakow,
G. Schierholz, R. Sommer and U. J. Wiese, Nucl. Phys.
{\bf B334} (1990) 527.
\bibitem{schi92} M. G\"ockeler, R. Horsley, P. Rakow, G. Schierholz
and R. Sommer, Nucl. Phys. {\bf B371} (1992) 713.
\bibitem{azco92} V. Azcoiti, G. Di Carlo and A. F. Grillo, Int. J. Mod. Phys.
{\bf A8} (1993) 4235; Phys. Lett. {\bf B305} (1993) 275.
\bibitem{Kocic94}  A. Koci\'{c}, Nucl. Phys. {\bf B} (Proc. Suppl.) {\bf 34}
(1994) 123.
\bibitem{zaragoza_new} V.~Azcoiti, G.~Di~Carlo, A.~Galante, A.~F.~Grillo,
V.~Laliena, C.~Piedrafita, to be published in {\em Proceedings of  the
International Symposium on Lattice  Field Theory, Melbourne, Australia,
11--15 July 1995}, to  appear in Nucl. Phys. {\bf B} (Proc. Suppl.),
hep-lat/9509037; \\
 V.~Azcoiti, G.~Di~Carlo, A.~Galante, A.~F.~Grillo,
V.~Laliena, C.~Piedrafita, Phys. Lett. {\bf B353} (1995) 279.
\bibitem{juel94} M. G\"ockeler, R.~Horsley, V.~Linke, P.~E.~L.~Rakow,
G.~Schierholz and H.~St\"uben, Nucl Phys. {\bf B} (Proc. Suppl.) {\bf 42}
(1995) 660.
\bibitem{horo91} A. M. Horowitz, Phys. Rev. {\bf D43} (1991) R2461.
\bibitem{rakow91} P. E. L. Rakow, Nucl. Phys. {\bf B356} (1991) 27.
\bibitem{Bit89}
K. Bitar, A. D. Kennedy, R. Horsley, S. Meyer and P. Rossi,
Nucl. Phys. {\bf B313} (1989) 348.
\bibitem{brezin76}
E. Brezin, J. C. Le Guillou and J. Zinn-Justin, in {\it Phase
Transitions and Critical Phenomena}, Vol. 6, p. 125,
eds. C. Domb and M. S. Green (Academic Press, London, 1976).
\bibitem{schier91a} G. Schierholz, Nucl. Phys. {\bf B} (Proc. Suppl.) {\bf 20}
(1991) 623.
\bibitem{alikhan92} A. Ali Khan, Nucl. Phys. {\bf B} (Proc. Suppl.) {\bf 30}
(1993) 733.
\bibitem{goeckeler91} M. G\"ockeler, Nucl. Phys. {\bf B} (Proc. Suppl.) {\bf
20}
(1991) 642.
\bibitem{horsley91} R. Horsley, M. G\"ockeler, E. Laermann, P. Rakow, G.
Schierholz, R. Sommer and U.-J. Wiese, Nucl. Phys. {\bf B} (Proc. Suppl.) {\bf
  20} (1991) 639.
\bibitem{lusch90} M. L\"uscher, Nucl. Phys. {\bf B341} (1990) 341.
\bibitem{schi90b} M. G\"ockeler, R. Horsley, E. Laermann, U.-J. Wiese, P. E. L.
Rakow, G. Schierholz and R. Sommer, Phys. Lett. {\bf B251} (1990), 567; {\em
  erratum} ibid. {\bf B256} (1991) 562.
\bibitem{diss}
A.~Ali~Khan, {\em PhD Thesis} (in German),
DESY Internal  Report T--94--02 (1994).
\bibitem{muller56} D. E. M\"uller, {\em A method for solving algebraic
equations
using an automatic computer}, Mathematical Tables and Aids to Computation {\bf
  10} (1956) 208; \\
B. Leavenworth, {\em Algorithm 25: Real zeros of an arbitrary function},
Communications of the ACM {\bf 3} (1960) 602.
\bibitem{kogut89} E. Dagotto, A. Koci\'{c} and J. Kogut, Phys. Lett. {\bf B232}
(1989) 235.
%
\bibitem{NJL} A. Ali Khan, M. G\"ockeler, R. Horsley, P.~E.~L. Rakow,
G. Schierholz
and H. St\"uben,  Phys. Rev. {\bf D51} (1995) 3751.
\bibitem{golterman86} M. F. L. Golterman, Nucl. Phys. {\bf B273} (1986) 663.
\end{thebibliography}
\end{document}